\documentclass[11pt,english]{article}

\usepackage[utf8]{inputenc} 

\usepackage{dsfont}

\usepackage[left=1in,right=1in,top=1in,bottom=1in]{geometry}
\usepackage{authblk}
\usepackage{enumitem}
\usepackage{ulem} 
\usepackage{empheq}

\usepackage{graphicx}
\usepackage{subcaption}
\usepackage{float}
\usepackage{wrapfig} 

\usepackage{xcolor}
\usepackage{color}
\definecolor{darkgreen}{rgb}{0,0.5,0}
\definecolor{darkblue}{rgb}{0,0,0.6}
\definecolor{purple}{rgb}{0.4,.2,0.7}

\usepackage{amsmath}
\usepackage{amssymb}

\usepackage{tensor} 
\usepackage{slashed}
\usepackage{mathtools}
\usepackage{leftidx}
\usepackage{esint}
\usepackage{amsfonts,amsthm,bm}

\usepackage{physics}
\usepackage{feynmf}
\usepackage{braket}

\usepackage{prettyref}
\usepackage[colorlinks=true,citecolor=darkgreen,linkcolor=purple,urlcolor=purple]{hyperref}

\usepackage[numbers,sort&compress]{natbib}

\newcommand{\nn}{\nonumber}

\numberwithin{equation}{section}
\numberwithin{figure}{section}
\numberwithin{table}{section}

\begin{document}

\title{\LARGE\textsc{Black Hole Horizon Edge Partition Functions}}  


\author[a]{\vskip1cm \normalsize Manvir Grewal}
\affil[a]{ \it  Center for Theoretical Physics, Columbia University, New York, NY 10027, USA}
\author[b,c]{\normalsize Y.T.\ Albert Law}
\affil[b]{ \it \normalsize	 Center for the Fundamental Laws of Nature, Harvard University, Cambridge, MA 02138, USA}
\affil[c]{ \it \normalsize	 Black Hole Initiative, Harvard University, Cambridge, MA 02138, USA}
\author[a]{Klaas Parmentier}
\date{}
\maketitle

\begin{center}
	\vskip-10mm
	{\footnotesize \href{mailto: mg3978@columbia.edu}{ manvir.grewal@columbia.edu}\, ,\;  \href{mailto:ylaw1@g.harvard.edu}{ylaw1@g.harvard.edu}\, ,\;  \href{mailto:k.parmentier@columbia.edu}{k.parmentier@columbia.edu}}
\end{center}

\vskip10mm

\thispagestyle{empty}


    

\begin{abstract}
We extend a formula for 1-loop black hole determinants by Denef, Hartnoll, and Sachdev (DHS) to spinning fields on any $(d+1)$-dimensional static spherically symmetric black hole. By carefully analyzing the regularity condition imposed on the Euclidean eigenfunctions, we reveal an unambiguous bulk-edge split in the 1-loop Euclidean partition function for tensor fields of arbitrary integer spin: the bulk part captures the ``renormalized" thermal canonical partition function recently discussed in \cite{Law:2022zdq}; the edge part is related to quasinormal modes (QNMs) that fail to analytically continue to a subset of Euclidean modes with enhanced fall-offs near the origin. Since the edge part takes the form of a path integral on $S^{d-1}$, this suggests that these are associated with degrees of freedom living on the bifurcation surface in the Lorentzian two-sided black hole geometry. For massive higher spin on static BTZ and massive vector on Nariai black holes, we find that the edge partition function is related to the QNMs with lowest overtone numbers.

\end{abstract}

\newpage

\tableofcontents


\newpage

\section{Introduction}

Ever since the seminal work \cite{Gibbons:1976ue}, the Euclidean gravitational path integral has been a prominent tool that has led to tremendous progress in thermodynamic and entanglement aspects of quantum black holes. In some cases, one finds exact agreement with microscopic calculations in string theory or holographic CFTs, even beyond the leading order in $G_N$ \cite{Banerjee:2010qc, Banerjee:2011jp, Sen:2012dw, Sen:2012kpz, Sen:2014aja}.

Operationally, one starts with a formal path integral integrating over all metrics and matter fields, and then expands $g= g^* +\delta g, \Phi = \Phi^* +\delta \phi$ around the saddle points $(g^*,\Phi^*)$
\begin{align}
	Z = \int \mathcal{D} g\, \mathcal{D}\Phi \, e^{-S[g,\Phi]} \approx \sum_{g^*,\Phi^*} e^{-S[g^*,\Phi^*]} Z_\text{1-loop} \left[g^*,\Phi^*\right] \left(1+\cdots \right) \; .
\end{align}
Many aspects of such a formal object remain to be understood. For example, what exactly should we sum over in $\sum_{g^*,\Phi^*}$? Another well-known confusion is that in the gravity sector\footnote{In general, for any massless fields with spin $s\geq 2$ there are finite number of modes with a wrong sign of kinetic term, which has been demonstrated explicitly in \cite{Law:2020cpj} for the case of massless higher spin fields on a sphere.} there is a conformal mode that renders the gravitational action unbounded from below \cite{Gibbons:1978ac}. 

This paper is a continuation of \cite{Law:2022zdq}, concerning the 1-loop contributions from matter fields and the graviton around a $(d+1)$-dimensional static spherically symmetric black hole background. In Euclidean signature, this means we always have $U(1)\times SO(d)$ symmetry, associated with the thermal circle and the codimension-2 sphere, as part of the isometries. In \cite{Law:2022zdq}, we considered the 1-loop Euclidean path integral for a real scalar on such a background, which takes the form of a functional determinant
\begin{align}\label{intro:eucs0}
	Z_\text{PI}(m^2)=\int\mathcal{D}\phi \, e^{-\frac{1}{2}\int \left(\nabla \phi \right)^2 + m^2 \phi^2}=\frac{1}{\det \left(-\nabla^2+m^2 \right)^{1/2}} \; .
\end{align}
Our key result in \cite{Law:2022zdq} is that \eqref{intro:eucs0} has a canonical interpretation through the relation
\begin{align}\label{intro:s0pican}
    Z_\text{PI}=\widetilde{Z}_\text{bulk} \; , \qquad \widetilde{Z}_\text{bulk} \equiv \frac{Z_\text{bulk}}{Z^\text{Rin}_\text{bulk}}   \qquad \qquad \text{(scalar)} \; .
\end{align}
Here $Z_\text{bulk}\equiv \Tr \, e^{-\beta_H \hat{H}}$ is the formal thermal canonical partition function at the inverse black hole temperature $\beta_H$ for the scalar living outside the horizon, while $Z^\text{Rin}_\text{bulk}$ is analogously defined but on a Rindler-like wedge at the inverse temperature $\beta_H$. As explained in \cite{Law:2022zdq} and briefly reviewed in Section \ref{sec:scattering}, while these traces $\Tr$ are ill-defined, their ratios can be unambiguously defined. Explicitly, the ``renormalized" partition function is given by the formula
\begin{align}\label{introeq:dhs}
    \log \widetilde{Z}_\text{bulk} = \int_0^\infty \frac{dt}{2t} \frac{1+e^{-2\pi t/\beta_H}}{1-e^{-2\pi t/\beta_H}} \chi_\text{QNM}(t) \; , \qquad  \chi_\text{QNM}(t) \equiv \sum_z N_z \, e^{-i z t}\;.
\end{align}
Here $\chi_\text{QNM}(t)$ is a ``quasinormal mode (QNM) character" defined as a sum over the QNM spectrum, with $z$ the frequencies of the QNMs and $N_z$ their degeneracies. The relation \eqref{intro:s0pican} has been verified in \cite{Law:2022zdq} for the case of scalars on static BTZ, Nariai, and the de Sitter static patch. 

In this work, we extend these considerations to arbitrary spinning fields. While we will focus on massive fields, since the 1-loop path integrals for massless gauge fields are given by ratios of determinants of differential operators, we can simply put together the massive results taking the masses to specific values in order to obtain the massless results.\footnote{For compact spaces there could be new subtleties for massless fields, such as residual group volume or Polchinski's phase coming from Wick-rotating the conformal modes \cite{Law:2020cpj,Gibbons:1978ac,Polchinski:1988ua}. These are contributions from a finite number of modes and do not affect the general consideration of this paper.}

As demonstrated by our explicit examples of massive higher spin (HS) fields on static BTZ and massive vectors on Nariai, one can still define $\widetilde{Z}_\text{bulk}$ as a formal ratio like \eqref{intro:s0pican}, which continues to be given explicitly by the formula \eqref{introeq:dhs}. However, it turns out that $Z_\text{PI}\neq \widetilde{Z}_\text{bulk}$ for any spin $s\geq 1$. In fact, for massive symmetric tensor \cite{Anninos:2020hfj,Sun:2020ame} and $p$-form fields \cite{David:2021wrw} with arbitrary spins on a round sphere $S^{d+1}$ or $EAdS_{d+1}$, it was observed that their Euclidean path integrals could be brought into the form
\begin{align}\label{introeq:ZPI}
    Z_\text{PI} = \frac{\widetilde{Z}_\text{bulk} }{Z_\text{edge}} \; .
\end{align}
Our goal is to provide an explanation for this bulk-edge split and systematically characterize the edge part $Z_\text{edge}$ for general higher spin fields on any static black hole background. 

To arrive at \eqref{introeq:ZPI}, we first note that the formula \eqref{introeq:dhs} is equivalent to a formula derived by Denef, Hartnoll, and Sachdev (DHS) \cite{Denef:2009kn} for the scalar Euclidean path integral \eqref{intro:eucs0}. The DHS derivation was based on the analytic properties of $Z_\text{PI}(m^2)$ as a function on the complex $m^2$-plane, and the fact that {\it any} QNM would Wick-rotate to a {\it regular} Euclidean mode at the correct (complex) value of $m^2$. 

As explained in Section \ref{sec:dhsspin}, the key subtlety for spinning fields is associated with the regularity condition (i.e. smoothness and single-valuedness around the Euclidean time direction) imposed on the field configurations included in the Euclidean path integration. While this condition seems innocuous (and is naturally assumed in any calculation of 1-loop determinants in the literature), a careful analysis reveals that it has non-trivial consequences in the DHS derivation, as already pointed out in the context of spin-2 fields on a BTZ background in \cite{Castro:2017mfj}.\footnote{In \cite{Castro:2017mfj}, the analysis was phrased in terms of (local) square integrability at the origin, which is implied if the functions are regular at the origin.} As we will see, the regularity condition for spinning fields creates an obstruction for some QNMs to Wick-rotate to a subset of regular Euclidean modes and eventually leads to the form \eqref{introeq:ZPI}, with the edge part explicitly given by 
\begin{align}\label{introeq:generaledge}
     \log Z_\text{edge} =\int_0^\infty \frac{dt}{2t} \sum_{k=-(s-1)}^{s-1}\sum_{z_{e,k}} e^{-\left( \frac{2\pi}{\beta_H} |k|+i z_{e,k} \right)t} 
\end{align}
for a massive spin-$s$ field. Here, $z_{e,k}=z_{e,k}(m^2)$ are the frequencies of those QNMs which fail to Wick-rotate to a regular Euclidean mode with $U(1)$ quantum number $|k|$ for any complex value of mass $m^2$. The sum $\sum_{z_{e,k}}$ receives contributions from $SO(d)$ representations of spin $0,1,\dots ,s-1$. Since $Z_\text{edge}$ is characterized based on the regularity condition near the origin, it is natural to associate these $SO(d)$ degrees of freedom as living on the bifurcation surface $S^{d-1}$ in the Lorentzian signature, thus justifying the terminology ``edge". In Sections \ref{sec:btzhs} and \ref{sec:nariai}, we work out the explicit form of \eqref{introeq:generaledge} for massive HS on static BTZ and massive vector on Nariai. In combination with \eqref{introeq:dhs}, we then find exact agreement with $Z_\text{PI}$ as in \eqref{introeq:ZPI}.



While in this work we do not have a canonical interpretation for $Z_\text{edge}$, the structure \eqref{introeq:ZPI} is generally expected from studies of entanglement entropy in gauge theories and gravity. We will comment more on this as we conclude in Section \ref{sec:discuss}.

As mentioned earlier on, the general form \eqref{introeq:ZPI} for higher spin fields on a sphere was first observed in \cite{Anninos:2020hfj}, where, however, the precise $SO(d)$ contents for $Z_\text{edge}$ were somewhat obscure. To clarify those, one could follow the same procedure of checking the Euclidean continuation of QNMs demonstrated in our explicit examples in Sections \ref{sec:btzhs} and \ref{sec:nariai}, and put $Z_\text{edge}$ into the form \eqref{introeq:generaledge}. However, it turns out that there exists yet another way to work out the precise $SO(d)$ contents for $Z_\text{edge}$, by exploiting powerful methods from representation theory. This will be explained in an upcoming work \cite{branching}.

\paragraph{Plan of the paper}

We review the DHS formula for scalars and its Lorentzian interpretation in Section \ref{sec:char deriv}. In Section \ref{sec:dhsspin}, we examine the Euclidean regularity condition and generalize the DHS arguments to arbitrary spinning fields. In Sections \ref{sec:btzhs} and \ref{sec:nariai} we work out the explicit examples of massive HS on static BTZ and massive vector on Nariai respectively. We collect some helpful basic facts for scalar and vector spherical harmonics in Appendix \ref{app:harmonics}. Appendices \ref{app:rindler}-\ref{app:btz} contain technical calculations that are useful in our analysis.



\section{Comments on the Denef-Hartnoll-Sachdev formula}\label{sec:char deriv}

\subsection{Review of the Denef-Hartnoll-Sachdev argument for scalars}

The following discussion applies to arbitrary ($d+1$)-dimensional static spherically symmetric backgrounds:
\begin{align}\label{sym metric}
	ds^2=-F(r)dt^2+\frac{dr^2}{F(r)}+r^2 d\Omega_{d-1}^2\, .
\end{align}
Here $d\Omega_{d-1}^2$ is the metric on the unit $S^{d-1}$. There is a horizon at $r=r_H$ if $F(r_H)=0$, with inverse Hawking temperature $\beta_H =\frac{1}{T_H}=\frac{4\pi}{F'(r_H)}$. Wick-rotating $t=-i t_E$ in \eqref{sym metric} and making it periodic
\begin{align}\label{E sym metric}
	ds^2 \to ds_E^2=F(r)d t_E^2+\frac{dr^2}{F(r)}+r^2 d\Omega_{d-1}^2\; ,\qquad t_E \simeq t_E +\beta_H \, ,
\end{align}
we obtain a smooth geometry that arises as a saddle point in the Euclidean gravitational path integral. The above analytic continuation maps the horizon at $r=r_H$ to the origin, near which we can make a change of variables
\begin{align}
	\rho^2 =\frac{4}{F'(r_H)}(r-r_H)\; ,\quad \varphi = \frac{2\pi}{\beta_H}t_E\; ,
\end{align}
so that the near-horizon spacetime takes the product form
\begin{align}\label{near hor metric}
	ds^2\approx d\rho^2 +\rho^2d\varphi^2+r_H^2\, d\Omega_{d-1}^2=du d\bar u +r_H^2\, d\Omega_{d-1}^2\; .
\end{align}
In the last equality we have introduced the complex coordinates
\begin{align}\label{eq:Eucuub}
    u=\rho \, e^{-i\varphi} \; ,\qquad \bar{u}=\rho \, e^{i\varphi}\; .
\end{align}
At 1-loop, corrections to the gravitational path integral are given by integrating quadratic fluctuations of matter fields (including the graviton) living on \eqref{E sym metric}. For instance, the 1-loop contribution of a real scalar $\phi$ with mass $m^2$ is given by
\begin{align}\label{eq:scalar PI}
	Z_\text{PI}(m^2)=\int\mathcal{D}\phi \, e^{-\frac{1}{2}\int \left(\nabla \phi \right)^2 + m^2 \phi^2}=\frac{1}{\det \left(-\nabla^2+m^2 \right)^{1/2}} \; .
\end{align}

\paragraph{Regularity condition}

We demand the functions in the functional integration \eqref{eq:scalar PI} to be smooth at the origin $\rho=0$ and single-valued in the Euclidean time direction, i.e. they should be regular functions. This means that $\phi$ has a Taylor expansion in the complex coordinates $u,\bar{u}$ near the origin. More precisely, a mode with thermal frequency $k$ has the following $\rho\to 0$ behavior:
\begin{align}\label{quan asym}
	\phi_k \sim \rho^{|k|} e^{-ik\varphi} =
	\begin{cases}
		u^k \; ,\quad &k\geq 0\\
		\bar{u}^{-k} \; ,\quad &k\leq 0
	\end{cases}\; .
\end{align}
As part of the definition of the path integral, $\phi$ is typically required to satisfy other boundary conditions (e.g. standard or alternate boundary condition in asymptotically $AdS$ black holes).

The idea of \cite{Denef:2009kn} is that we assume the functional determinant \eqref{eq:scalar PI} to be a meromorphic function on the complex $m^2$-plane, and try to match its poles and zeros. $Z_\text{PI}(m^2)$ has no zero, and hits a pole whenever
\begin{align}\label{s0 eom}
	(-\nabla^2+m^2)\, \phi=0 \; .
\end{align}
Solving this equation near $\rho=0$, we deduce the near-origin behavior
\begin{align}\label{gen asm}
	\phi \sim \rho^{\mp i\frac{z}{2\pi T_H}}e^{-\frac{z}{2\pi T_H}\varphi} =\rho^{\mp i\frac{z}{2\pi T_H}}e^{-zt_E},
\end{align}
where $z=z(m^2)$ is a function of $m^2$. If we Wick rotate \eqref{gen asm} back to real time, the  $\rho\to 0$ behavior becomes the near-horizon behavior
\begin{align}
	\phi \sim \rho^{\mp i\frac{z}{2\pi T_H}}e^{-izt}=e^{-iz(t\pm x)},\quad x\equiv \frac{\ln \rho}{2\pi T_H}\; .
\end{align}
This is the boundary condition satisfied by (anti-)QNMs purely approaching (leaving) the horizon. Therefore, for {\it physical} $m^2$, $z$ is a QNM or anti-QNM frequency.

Now, for $m^2$ to be a pole of $Z_{\text{PI}}(m^2)$, we need the Euclidean solution $\phi$ to be regular at the origin. We can see from \eqref{gen asm} that for {\it generic} $m^2$ this will not be the case. However, by varying $m^2$ (and thus $z(m^2)$) over the complex plane, we encounter a regular solution of \eqref{s0 eom} every time that \eqref{gen asm} matches onto either branch of \eqref{quan asym}. The $\mp$ branch in \eqref{gen asm} can only be matched onto the $k>0$ ($k<0$) branch in \eqref{quan asym}, while either can match onto the $k=0$ mode. Therefore, we conclude that\footnote{Generally there is a holomorphic function $e^{P(m^2)}$ multiplying \eqref{eq:qnm det formula}, which is related to the UV-divergences (including the logarithmic divergence $d+1$ is even) of \eqref{eq:qnm det formula} and can be determined by comparing $m^2\to \infty$ asymptotics of \eqref{eq:qnm det formula} and the heat kernel coefficients \cite{Denef:2009kn}. We proceed formally neglecting this issue, and will provide a rigorous regularization when we discuss explicit examples.}
\begin{align}\label{eq:qnm det formula}
	\frac{1}{\det \left(-\nabla^2+m^2 \right)}
	=&\prod_{z,\bar{z}}\prod_{k=-\infty}^\infty \left( |k|+ \frac{iz}{2\pi T_H}\right)^{-N_{z}/2}\left( |k|- \frac{i\bar{z}}{2\pi T_H}\right)^{-N_{\bar z}/2} \; .
\end{align}
Here $\bar z$ are anti-QNM frequencies. When the theory is PT-symmetric, $\bar z$ can be taken to be the complex conjugate of $z$. Alternatively, we observe that since the Lorentzian equation of motion is invariant under $t\to -t$, for a QNM with frequency $z$, there is an anti-QNM with frequency $-z$. Therefore, we can replace $\bar z \to -z$ in \eqref{eq:qnm det formula}, and we have simply
\begin{align}\label{eq:qnmPI}
    Z_\text{PI}(m^2)= \prod_z \prod_{k=-\infty}^\infty \left( |k|+ \frac{iz}{2\pi T_H}\right)^{-N_z/2} \; .
\end{align}
We will focus on this case from now on. Using $\log x = \int_0^\infty \frac{dt}{t} e^{-x t}$ (ignoring the issue of UV-divergence), we can formally write \cite{Law:2022zdq}
\begin{align}\label{eq:dhsformula}
\log Z_\text{PI} = \int_0^\infty \frac{dt}{2t}\sum_{z} \sum_{k=-\infty }^\infty N_z \, e^{-\left( |k|+ \frac{iz}{2\pi T_H}\right)t} =\int_0^\infty \frac{dt}{2t}\frac{1+e^{-2\pi t/\beta_H}}{1-e^{-2\pi t/\beta_H}}\chi_\text{QNM}(t)\; .
\end{align}
In the second equality we performed the sum over $k$, scaled $t\to 2\pi t/\beta_H$, and expressed in terms of the ``QNM character"
\begin{align}\label{eq:qnm char}
\chi_\text{QNM}(t)\equiv \sum_{z} \, N_z \, e^{-izt}\; .
\end{align}

\subsection{Black hole scattering and the renormalized partition function}\label{sec:scattering}

The main result of \cite{Law:2022zdq} is a Lorentzian calculation that reproduces the 1-loop Euclidean path integral as computed by the DHS formula \eqref{eq:dhsformula}, which we review in this section. We refer the reader to \cite{Law:2022zdq} for a more detailed discussion. For concreteness we focus on the case of asymptotically AdS black holes for $d\geq 3$. 

\paragraph{Black hole scattering}

To start with, we separate
\begin{align}\label{eq:normalmodes}
	\phi_{\omega l} (t,r,\Omega)= e^{-i \omega t}\,\frac{\psi_l (r)}{r^\frac{d-1}{2} } \, Y_l(\Omega)\; .
\end{align}
For every integer $l\geq 0$, $Y_l$ are the $(d-1)$-dimensional spherical harmonics. Making use of this ansatz and the tortoise coordinate $x \equiv \int^r_\infty\frac{dr'}{F(r')}$, the Klein-Gordon equation $\left( -\nabla^2+m^2\right)\phi=0$ on the background \eqref{sym metric} is recast into a 1D Schr\"odinger form for each $l$:
\begin{align}\label{eq:eff Sch}
	\left( -\partial_{x}^2 +V_l(x) \right) \psi_l (x) =\omega^2 \psi_l (x)\; ,
\end{align}
with the effective potential 
\begin{align}\label{eq:potential}
	V_l(x) = F(r) \left[ \frac{d-1}{2r^{\frac{d-1}{2}}} \partial_r \left(r^{\frac{d-3}{2}} F(r) \right) +\left( \frac{l(l+d-2)}{r^2}+m^2\right)  \right] \, .
\end{align}
In the near-horizon regime ($x\to -\infty$), the normalizable solution to \eqref{eq:eff Sch} satisfying the standard boundary condition at infinity ($x=0$) takes the asymptotic form
\begin{align}\label{eq:nearhor}
	\psi_l (x\to -\infty ) \sim A_l^\text{out}(\omega)\, e^{-i\omega x} + A_l^\text{in}(\omega) \, e^{i\omega x}  \; .
\end{align}
Here by ``in" (``out") we mean the waves travel away from (towards) the horizon, as opposed to the common terminology in studies of QNMs. For real $\omega$, $A_l^\text{in}(\omega)={A_l^\text{out}}^*(\omega)$, and the ratio
\begin{align}\label{eq:s matrix cos}
	\mathcal{S}_l (\omega) = \frac{A_l^\text{out}(\omega) }{A_l^\text{in}(\omega) } \equiv e^{2i\theta_l(\omega)}
\end{align}
is a pure phase, or a rank-1 unitary S-matrix.

\paragraph{The renormalized partition function}

A naive Lorentzian calculation to be compared with the 1-loop Euclidean path integral \eqref{eq:scalar PI}, would be that of the ideal gas canonical partition function for the scalar field living on the background \eqref{sym metric}:
\begin{align}\label{eq:freeen}
    \log Z_\text{bulk} \equiv \log \Tr \, e^{-\beta_H \hat{H}}= \int_0^\infty d \omega \, \rho(\omega) \log \left(e^{\beta_H \omega/2}-e^{-\beta_H \omega/2} \right) \; .
\end{align}
Here $\rho(\omega) = \sum_l D_l^d \rho_l(\omega)$ is the total single-particle density of states (DOS). As it is, \eqref{eq:freeen} is pathological: for every $SO(d)$ angular momenta $l\geq 0$, there is a continuum of normal modes in any small interval $\Delta \omega$, and thus $\rho_l(\omega)$ is strictly infinite . This infinity is distinct from the usual UV-divergences coming from integrating over all $\omega>0$ and summing over all $l\geq 0$.

The key realization of \cite{Law:2022zdq} is that the non-trivial information about the spacetime and the scalar field encoded in the potential \eqref{eq:potential} can be extracted by comparing the scattering problem \eqref{eq:eff Sch} to a reference problem with potential $\bar V_l(x)$. The difference of $\rho_l(\omega)$ from the reference $\bar \rho_l (\omega)$ is a completely finite quantity, related to the scattering matrices \eqref{eq:s matrix cos}:
\begin{align}\label{eq:rhochange}
    \Delta\rho_l (\omega)=\rho_l (\omega)-\bar\rho_l (\omega) =
	\frac{1}{2\pi i}\partial_\omega \left(\log \mathcal{S}_l(\omega) - \log \bar{\mathcal{S}}_l(\omega)\right) \; .
\end{align}
Here the difference in the first equality is understood in a limiting sense explained in \cite{Law:2022zdq}. Therefore, instead of \eqref{eq:freeen}, a class of better defined objects are given by {\it differences} of free energies:
\begin{align}\label{eq:freeendiff}
     \log Z_\text{bulk} -\log \bar Z_\text{bulk}= \int_0^\infty d \omega \, \Delta\rho(\omega) \log \left(e^{\beta_H \omega/2}-e^{-\beta_H \omega/2} \right)
\end{align}
where
\begin{align}\label{eq:fullrhochange}
	\Delta \rho(\omega) = \sum_{l=0}^\infty D_l^d \, \Delta \rho_l(\omega)=
	\frac{1}{2\pi i}\partial_\omega \sum_{l=0}^\infty D_l^d \left(\log \mathcal{S}_l(\omega) - \log \bar{\mathcal{S}}_l(\omega)\right) \; .
\end{align}
Quantities like \eqref{eq:freeendiff} are still UV-divergent due to the integration over all $\omega$ and the sum over all $l\geq 0$, but these are the usual divergences that are absorbed into the renormalization of the cosmological constant, Newton's constant and curvature couplings once we couple our theory to gravity. 

A priori, there is no canonical choice of the reference scattering problem. For example, one could consider the reference with the minimal potential $\bar V_l (x) =0$. Any choice of $\bar V_l (x) $ would lead (after UV-regularization) to a finite ``renormalized" free energy \eqref{eq:freeendiff}. By working out the examples for scalars on static BTZ, Nariai and static patch in de Sitter, \cite{Law:2022zdq} observes that choosing $\bar Z_\text{bulk}$ to be that on a Rindler-like wedge at the inverse black hole temperature $\beta_H$, with the associated scattering problem 
\begin{align}\label{eq:nearhorscatt}
    \left[ -\partial_{x}^2 + V^\text{Rin} (\beta_H,x) \right] \psi(x) = \omega^2 \psi(x) \; , \qquad V^\text{Rin} (\beta, x)\equiv \left(\frac{4\pi}{\beta} \right)^2 e^{\frac{4\pi}{\beta}x} \; ,
\end{align}
the renormalized free energy equals the 1-loop Euclidean partition function: 
\begin{align}
    \widetilde{Z}_\text{bulk} = Z_\text{PI}\;, \qquad \widetilde{Z}_\text{bulk} \equiv \frac{Z_\text{bulk}}{Z^\text{Rin}_\text{bulk}(\beta_H)} \qquad \text{(scalar)} \; .
\end{align}
In particular, from all the examples, one finds that the S-matrix for the original problem always takes the form $\mathcal{S}_l(\omega)=\mathcal{S}^\text{QNM}_l(\omega)\mathcal{S}^\text{Rin} (\beta_H,\omega)$, where $\mathcal{S}^\text{QNM}_l(\omega)$ contains QNM frequencies as poles and anti-QNM frequencies as zeros, and 
\begin{align}\label{eq:RindlerS}
    \mathcal{S}^\text{Rin} (\beta,\omega) = \frac{\Gamma
   \left(\frac{i \beta  \omega }{2 \pi }\right)}{\Gamma
   \left(-\frac{i \beta  \omega }{2 \pi }\right)} \; .
\end{align}
is the scattering matrix for the Rindler problem \eqref{eq:nearhorscatt}, which has the Matsubara frequencies as zeros and poles. Therefore, choosing the Rindler problem \eqref{eq:nearhorscatt} as the reference, the renormalized DOS is the Fourier transform of the QNM character
\begin{align}\label{eq:renDOS}
    \Delta \rho(\omega) = \frac{1}{2\pi i} \sum_z N_z \left(\frac{1}{\omega+z}-\frac{1}{\omega-z}\right) =\int_0^\infty \frac{dt}{2\pi} \left(e^{i\omega t} +e^{-i\omega t}  \right) \chi_\text{QNM}(t) \; .
\end{align}
While in principle there could be a holomorphic part contributing to $\Delta \rho(\omega)$, in all the explicit examples $\Delta \rho(\omega)$ does not receive such a contribution and \eqref{eq:renDOS} gives the complete answer. Plugging \eqref{eq:renDOS} and performing the $\omega$-integral gives the DHS formula \eqref{eq:dhsformula}.

\paragraph{Generalization to higher spins}

To conclude this section, we note that the Lorentzian considerations above readily generalize to spinning fields. In Appendix \ref{app:rindler}, we explicitly solve the scattering problem for a massive higher spin (HS) field on the Rindler-like wedge and obtain the S-matrices. 

In Section \ref{sec:btzhs}, we study the example of massive higher spin on static BTZ. For any spin $s\geq 1$, it remains true that for each angular momentum $l\in\mathbb{Z}$, the S-matrix for the associated problem takes the product form $\mathcal{S}_l(\omega)=\mathcal{S}^\text{QNM}_l(\omega)\mathcal{S}^\text{Rin,(s)} (\beta_H,\omega)$, where $\mathcal{S}^\text{QNM}_l(\omega)$ contains QNM frequencies as poles and anti-QNM frequencies as zeros, and $\mathcal{S}^\text{Rin,(s)} (\beta_H,\omega)$ is the Rindler S-matrix \eqref{appeq:rinhsSmat} generalized to a spin-$s$ field. Therefore, choosing the reference to be the Rindler problem, the renormalized free energy is still given by the formula
\begin{align}\label{eq:DHS}
\log \widetilde{Z}_\text{bulk}  =\int_0^\infty \frac{dt}{2t}\frac{1+e^{-2\pi t/\beta_H}}{1-e^{-2\pi t/\beta_H}}\chi_\text{QNM}(t)\; ,
\end{align}
where the QNM character $\chi_\text{QNM}(t)$ is analogously defined as \eqref{eq:qnm char}. However, in contrast to the scalar case, this turns out {\it not} to be equal to the 1-loop Euclidean path integral. The latter needs to be modified by ``edge" corrections. We turn to this next.



\section{Edge partition functions for spinning fields}\label{sec:dhsspin}

In this section we extend the DHS formula to spinning fields. As we will see, the regularity condition in Euclidean signature is more subtle than its scalar counterpart; in certain sectors some components are required to have enhanced fall-offs near the origin. This eventually leads to a natural bulk-edge split for the Euclidean path integral.

\subsection{Spin-1}

As an illustration of the idea, we first consider a massive vector $A^\mu$ living on the background \eqref{sym metric}, with the 1-loop path integral
\begin{align}\label{eq:vector PI}
	Z_\text{PI}(m^2)=\int \mathcal{D}A \, e^{-\int \left(\frac{1}{4}F_{\mu\nu}F^{\mu\nu} +\frac{m^2}{2} A_\mu A^\mu\right)}=\det \left(-\nabla_{(1)}^2+m^2 \right)^{-1/2} \, .
\end{align}
We denote $-\nabla_{(1)}^2$ as the Laplacian acting on transverse vector fields. On compact spaces such as a sphere, there will be an extra correction due to a normalizable constant scalar mode \cite{Anninos:2020hfj,Law:2020cpj}. The inclusion of this mode is essential for consistency with locality and unitarity \cite{Donnelly:2013tia}. Here we neglect such a contribution to keep the argument as simple as possible. Such a subtlety will matter when we study the example of Nariai spacetime in Section \ref{sec:nariai}.

\paragraph{Regularity condition and the analyticity argument}


Similar to the scalar case, we demand the vector fields $A_\mu$ in the functional integration \eqref{eq:vector PI} to be smooth at the origin $\rho =0$ and single-valued in the Euclidean time direction, i.e. they are regular vector fields. Once again, the most convenient way to assess regularity is to work with the complex coordinates $u=\rho\, e^{-i\varphi}$ and $\bar{u}=\rho\, e^{i\varphi}$. At the origin these are well-defined, unlike the polar coordinates $(\rho,\varphi)$. The components $(A_u,A_{\bar{u}})$ are related to $(A_\rho,A_{\varphi})$ through 
\begin{align}\label{original relate}
	A_\rho = e^{-i\varphi}A_u+e^{i\varphi}A_{\bar{u}}\quad ,\qquad 
	A_\varphi= -i\rho \left(e^{-i\varphi}A_u-e^{i\varphi}A_{\bar{u}}\right) \; .
\end{align}
A mode with $U(1)$ quantum number $k\in \mathbb{Z}$, takes the form
\begin{align}
	\left( A_\rho \, , A_\varphi \, ,A_i \right) \propto e^{-ik\varphi}\; ,
\end{align}
which from \eqref{original relate} implies that $A_u$, $A_{\bar u}$ and $A_i$ contain the factors
\begin{align}\label{eq:factors}
    A_u \propto e^{-i(k-1)\varphi}\;, \quad A_{\bar{u}} \propto e^{-i(k+1)\varphi}\;,\quad A_i \propto e^{-ik\varphi} \; .
\end{align}
The regularity condition boils down to requiring $(A_u,A_{\bar{u}},A_i)$ to have a Taylor series expansion in $u$ and $\bar{u}$ near the origin, which means that the leading term of the $\rho$-expansions of $A_u$, $A_{\bar u}$ and $A_i$ must combine with \eqref{eq:factors} to form non-negative powers of $u$ or $\bar u$. Explicitly, the result is 
\begin{itemize}
	\item for $k\geq 1$, 
	\begin{align}\label{eq:veck1}
		A_u \sim u^{k-1}\;, \quad A_{\bar{u}} \sim u^{k+1}\;,\quad A_i \sim u^k \; ;
	\end{align}
	
	\item for $k\leq -1$,
	\begin{align}\label{eq:veck-1}
		A_u \sim \bar{u}^{-k+1}\;, \quad A_{\bar{u}} \sim \bar{u}^{-k-1}\;,\quad A_i \sim \bar{u}^{-k} \; ;
	\end{align}
	
	\item for $k=0$,
	\begin{align}\label{eq:veck0}
		A_u \sim \bar{u}\;, \quad A_{\bar{u}} \sim u\;,\quad A_i \sim u^0 \; .
	\end{align}
\end{itemize}
Observe that when $k=0$, the fall-offs of $A_u$ and $A_{\bar u}$ do not follow the same pattern as the generic $|k|\geq 1$ sectors. Indeed, naively putting $k=0$ in \eqref{eq:veck1} or \eqref{eq:veck-1} would lead to a mode that diverges at the origin. This highlights the qualitative difference between spinning fields and scalars.

Now we repeat the DHS analyticity argument for the path integral \eqref{eq:vector PI} as a function on the complex $m^2$-plane. The functional determinant hits a zero whenever 
\begin{align}\label{s1 eom}
	(-\nabla_{(1)}^2+m^2)A_\mu=0
\end{align}
has a solution on the space of smooth vector fields. Here comes an important difference compared to the scalar case. Recall that in the latter case, {\it any} QNM would Wick rotate to a regular Euclidean mode with $U(1)$ quantum number $ k\in \mathbb{Z}_{\geq 0}$ as we vary $m^2$ so that
\begin{align}
    \frac{iz}{2\pi T_H}=-k \;.
\end{align}
This is not true for the massive vector: because of the enhanced fall-off \eqref{eq:veck0}, a {\it subset} of QNMs cannot be Wick-rotated to the $k=0$ sector. Similar comments apply to anti-QNMs. We will focus on PT-symmetric theories, where $\bar z$ can be taken to be $-z$.

\paragraph{Edge partition function}

Denoting  by $z_e$ the QNMs that {\it cannot} be Wick-rotated to the $k=0$ Euclidean modes due to the fall-off condition \eqref{eq:veck0}, we have a modified DHS formula: 
\begin{align}\label{BH vec det}
	Z_\text{PI}= \frac{\widetilde{Z}_\text{bulk}}{Z_\text{edge}}
\end{align}
where 
\begin{align}
	\log \widetilde{Z}_\text{bulk} = \int_0^\infty \frac{dt}{2t}\frac{1+e^{-2\pi t/\beta}}{1-e^{-2\pi t/\beta}}\chi_\text{QNM}(t) \; , \qquad \log Z_\text{edge} =\int_0^\infty \frac{dt}{2t} \sum_{z_e} e^{-i z_e t} \; .
\end{align}
Here $\chi_\text{QNM}(t)$ is the QNM character defined analogously as \eqref{eq:qnm char}. As discussed at the end of Section \ref{sec:char deriv}, the bulk part $\widetilde{Z}_\text{bulk}$ has an unambiguous meaning of a Rindler-renormalized thermal canonical partition function. Our analyticity argument reveals that the Euclidean path integral demands a division by the edge partition function $Z_\text{edge}$, which accounts for the fact that some QNMs cannot be Wick-rotated to $k=0$ Euclidean modes due to the fall-off condition \eqref{eq:veck0}. 

Note that the modes with frequency $z_e$ must have non-zero $A_u$ and $A_{\bar u}$ components, which must be constructed from $SO(d)$ scalars; therefore, $Z_\text{edge}$ can be thought of as a path integral of a scalar on $S^{d-1}$. Since our argument is based on the behavior of the vector field near the origin, it is natural to identify this $S^{d-1}$ with the bifurcation surface in the Lorentzian signature, thus justifying the terminology ``edge".


\subsection{Spin-2 and beyond}

The argument above readily generalizes to higher spin fields. For instance, for a symmetric spin-2 field $h_{\mu\nu}$, requiring $h_{uu},h_{u\bar u},h_{\bar u\bar u},h_{u i},h_{\bar u i},h_{ij}$ to have a Taylor expansion in $u,\bar u$ leads to the fall-offs near the origin:
\begin{itemize}
	\item for $k\geq 2$,
	\begin{align}
		h_{uu}\propto u^{k-2},\quad h_{u \bar u} \propto u^k ,\quad h_{\bar u \bar u} \propto u^{k+2},\quad h_{u i} \propto u^{k-1},\quad h_{\bar u i} \propto u^{k+1},\quad h_{i j} \propto u^{k} \; ;
	\end{align}

\item for $k=1$,
\begin{align}
	h_{uu}\propto \bar u,\quad h_{u \bar u} \propto u ,\quad h_{\bar u \bar u} \propto u^{3},\quad h_{u i} \propto u^{0},\quad h_{\bar u i} \propto u^{2},\quad h_{i j} \propto u \; ;
\end{align}

\item for $k=0$,
\begin{align}
	h_{uu}\propto \bar u^2,\quad h_{u \bar u} \propto u^0 ,\quad h_{\bar u \bar u} \propto u^{2},\quad h_{u i} \propto \bar u,\quad h_{\bar u i} \propto u,\quad h_{i j} \propto u^0 \; ;
\end{align}

\item for $k=-1$,
\begin{align}
	h_{uu}\propto \bar u^3,\quad h_{u \bar u} \propto \bar u ,\quad h_{\bar u \bar u} \propto u,\quad h_{u i} \propto \bar u^{2},\quad h_{\bar u i} \propto {\bar u}^{0},\quad h_{i j} \propto \bar u \; ;
\end{align}

	\item for $k\leq -2$,
\begin{align}
	h_{uu}\propto \bar u^{-k+2} ,\quad h_{u \bar u} \propto \bar u^{-k} ,\quad h_{\bar u \bar u} \propto \bar u^{-k-2},\quad h_{u i} \propto \bar u^{-k+1},\quad h_{\bar u i} \propto \bar u^{-k-1},\quad h_{i j} \propto \bar u^{-k} \; .
\end{align}
\end{itemize}
    
    
Observe the enhanced fall-off of $h_{uu}$ in the $k=0,1$ sector, that of $h_{\bar u \bar u}$ for $k=0,-1$, and those of $h_{u i},h_{\bar u i}$ for $k=0$. Note that $h_{uu},h_{\bar u \bar u}$ can only be constructed from $SO(d)$ scalars, while $h_{u i},h_{\bar u i}$ can be constructed from either $SO(d)$ scalars or vectors. 

The pattern goes on for tensor fields of arbitrary rank $s\geq1$. The component
\begin{align}
    \phi_{\underbrace{u\cdots u}_{a}\underbrace{\bar u\cdots \bar u}_{b}\underbrace{i_1\cdots i_c}_{c}} \; , \qquad a+b+c=s\;,
\end{align}
would have enhanced fall-offs for  $k=0,1,\dots , a-b-1$ if $a> b$, or $k=0,-1,\dots , a-b+1$ if $a< b$. Such a component can be constructed from $SO(d)$ representations of spin $0,1,\dots , c$. Repeating the analyticity argument, we then expect that for each fixed $|k|=0,1,\dots,s-1$, there will be a subset of (anti-)QNMs with frequencies $z_{e,k}$ ($\bar z_{e,k}$) that cannot be Wick-rotated to regular Euclidean modes of $U(1)$ quantum number $|k|$ ($-|k|$). For PT-symmetric theories where $\bar z_{e,k}$ can be taken to be $-z_{e,k}$, the edge partition function will take the general form
\begin{align}\label{eq:generaledge}
     \log Z_\text{edge} =\int_0^\infty \frac{dt}{2t} \sum_{k=-(s-1)}^{s-1}\sum_{z_{e,k}} e^{-\left( \frac{2\pi}{\beta_H} |k|+i z_{e,k} \right)t} \;.
\end{align}
In Sections \ref{sec:btzhs} and \ref{sec:nariai}, we will work out the explicit expressions of $Z_\text{edge}$ for massive HS fields on static BTZ and massive vector on Nariai, and check \eqref{BH vec det} against the full Euclidean path integrals obtained by direct derivations.


\section{Example: Massive higher spin on static BTZ}\label{sec:btzhs}

As our prime example, we consider massive higher spin (HS) fields living on the static BTZ background (setting $\ell_\text{AdS}=1$):
\begin{align}\label{eq:btzmetric}
    ds^2 = -\left(r^2-r_H^2 \right) dt^2 + \frac{dr^2}{r^2-r_H^2} +r^2 d\phi^2 = \frac{r_H^2}{\sinh^2 (r_H x)} \left( -dt^2 + dx^2 + \cosh^2 (r_H x) d\vartheta^2 \right)  \; .
\end{align}
In the second equality we have written in terms of the tortoise coordinate $r(x) = -r_H \coth(r_H x)$. We recall that $r_H\equiv M_\text{BH} =2\pi T_H$.

A spin-$s$ ($s\geq 1$) field of mass $m^2= (\Delta -s)(\Delta+s-2)$ living on such a background is described by either $(\mp)$ set of first-order equations \cite{Tyutin:1997yn}
\begin{align}\label{eq:btzhs1st}
    \epsilon\indices{_{\mu_1}^\alpha^\beta} \nabla_\alpha \phi_{\beta \mu_2 \cdots \mu_s}=\mp M \phi_{\mu_1 \mu_2 \cdots \mu_s} \;, \qquad M =\Delta -1\;.
\end{align}
In the current setting we are interested in the parity-invariant theory that includes both $\pm$ solutions. 

It turns out to be natural to study components with respect to the coordinates
\begin{align}\label{eq:new coor}
    y_\pm = e^{\mp r_H t} \sech (r_H x) \;,
\end{align}
in terms of which the metric becomes 
\begin{align}\label{eq:newbtzmetric}
    ds^2 &= \frac{1}{4(1- y_+ y_-)^2} \Big( y_-^2 dy_+^2 + 2(2-y_+y_-) dy_+dy_- + y_+^2 dy_-^2 \Big) + \frac{r_H^2}{1-y_+y_-} d\vartheta^2 \; .
\end{align}
Notice that near horizon $x\to -\infty$, $y_\pm \to 2 \, e^{r_H (x\mp t)}$ and
\begin{align}
    ds^2 \approx dy_+ dy_- +r_H^2 d\vartheta^2 \; .
\end{align}
Comparing this to \eqref{near hor metric}, we see that $y_+$ and $y_-$ are essentially the Lorentzian analogs for the complex coordinates $\bar u$ and $u$ in \eqref{eq:Eucuub} respectively. Thus, working with the components with respect to these coordinates make the comparison with Section \ref{sec:dhsspin} more direct.

Upon Wick-rotating $t\to-i t_E$ and identifying $t_E\sim t_E + \beta$ in \eqref{eq:btzmetric}, the resulting Euclidean BTZ (EBTZ) geometry is related to thermal $AdS_3$ ($TAdS_3$) by a large diffeomorphism. As a result, their path integrals are equal upon the modular transformation
\begin{align}\label{eq:mod trans}
	\tau \to -\frac{1}{\tau}\,\qquad \tau=2\pi i T_H \; .
\end{align}
This is expected to be true for any theories. Reproducing the $TAdS_3$ results (reviewed in Appendix \ref{app:tads}) hence serves as a consistency check for our method after we obtain $Z^\text{BTZ}_\text{PI}$ in Section \ref{sec:edgebtzpi}.

\subsection{Explicit solutions, scattering matrices, and quasinormal modes}

While the system \eqref{eq:btzhs1st} for massive HS fields on BTZ has been solved in for example \cite{datta_higher_2012}, we present a simpler version of this computation in Appendix \ref{app:btz}, where we work with components with respect to the coordinates \eqref{eq:newbtzmetric}.

\subsubsection{Massive scalar} 

We start with the simplest case of a massive scalar, whose normal mode functions will serve as the seed solutions for constructing those for general massive HS fields.

For a scalar with mass $m^2 = \Delta(\Delta-2)$, the normal mode solutions to the Klein-Gordon equation $\left( -\nabla^2+m^2\right)\phi=0$ are solved with the ansatz
\begin{align}\label{eq:btzs0an}
    \phi(t,x,\vartheta) =e^{-i\omega t+i l \vartheta} \sqrt{-\tanh (r_H x)}\, \psi^\text{Scalar}_{\omega l}(x) \;.
\end{align}
The normalizable solution satisfying the standard boundary condition is
\begin{align}\label{eq:btzscalarsol}
    \psi^\text{Scalar}_{\omega l}(x) = \frac{(\cosh\left(r_H x \right) )^{\frac{il}{ r_H}}\left(-\sinh \left(r_H x \right) \right)^{\Delta }}{\sqrt{-\tanh(r_H x)}} \, _2F_1\left(a_{\omega l} ,a_{-\omega l};\Delta ;-\sinh^2 \left(r_H x \right) \right) \;,
\end{align}
where
\begin{align}\label{eq:hyperpara}
    a_{\omega l}=\frac{\Delta}{2}+\frac{i  (-\omega+l )}{2 r_H}\;.
\end{align}
This solution has the near-horizon behavior
\begin{align}\label{eq:btzs0nh}
	 \psi^\text{Scalar}_{\omega l}(x\to -\infty) \propto  \frac{ \Gamma \left(\frac{i\omega }{r_H}\right) }{\Gamma
		\left(a_{-\omega, l} \right)\Gamma \left(a_{-\omega, -l} \right) } e^{-i  \omega x }+ \frac{ \Gamma \left(-\frac{i \omega }{r_H}\right) }{\Gamma
		\left(a_{\omega l} \right)\Gamma \left(a_{\omega, -l} \right)} e^{i  \omega x }\; .
\end{align}
The ratio of the incoming and outgoing coefficients defines a unitary S-matrix:
\begin{gather}
    \mathcal{S}_l (\omega) = \mathcal{S}_l^\text{BTZ} (\omega) \,\mathcal{S}^\text{Rin} \left(\frac{2\pi}{r_H},\omega\right) \label{eq:btzsmatrix}\nn\\
   \mathcal{S}_l^\text{BTZ} (\omega)=\mathcal{S}_l^{\text{BTZ},L} (\omega)\mathcal{S}_l^{\text{BTZ},R} (\omega)\; , \qquad  \mathcal{S}_l^{\text{BTZ},L} (\omega)=\mathcal{S}_{-l}^{\text{BTZ},R} (\omega) \equiv\frac{\Gamma
		\left(a_{\omega l} \right)}{\Gamma \left(a_{-\omega, -l} \right)} \; .
\end{gather}
Here $\mathcal{S}^\text{Rin} \left(\beta,\omega\right)$ is the Rindler S-matrix \eqref{eq:RindlerS}. The poles of $\mathcal{S}_l^{\text{BTZ}} (\omega)$ are the QNM frequencies 
\begin{align}\label{eq:btzs0qnm}
    z^L_{nl}= l - 2\pi T_H i (\Delta+2n)\; , \qquad z^R_{nl}=- l - 2\pi T_H i (\Delta+2n) \;,
\end{align}
while its zeros are the anti-QNM frequencies $-z^{L/R}_{nl}$ or $\left(z^{L/R}_{nl}\right)^*$. 

\subsubsection{Massive higher spin}

As explained in Appendix \ref{app:btz}, the incoming and outgoing behaviors for a normal mode solution to \eqref{eq:btzhs1st} are dominated by the components with all $+$- and $-$-indices respectively. We will focus on these and use the shorthand notation\footnote{In terms of the notation \eqref{appeq:hsnotation} in Appendix \ref{app:btz}, $\phi_{(+)}=\phi_{(s)(0)(0)}$ and $\phi_{(-)}=\phi_{(0)(s)(0)}$.}
\begin{align}\label{eq:notation}
    \phi_{(\pm)} \equiv \phi_{\underbrace{\pm\cdots \pm}_{s}} \; .
\end{align}
With our explicit calculations in Appendix \ref{app:btz}, we find the normal mode solutions to be 
\begin{equation}\label{eq:btzallpm}
	\phi^{(\mp)}_{(\pm)} = C^{(\mp)}_{\omega l,(\pm)}\, e^{ \pm s r_H t - i \omega t+i l\vartheta} (-\tanh(r_H x))^{\frac{1}{2}-s} \psi^\text{Scalar}_{\omega\pm i s r_H, l}(x) \; ,
\end{equation}
with $\psi^\text{Scalar}_{\omega l}(x)$ defined in \eqref{eq:btzscalarsol}. Here the superscript $(\mp)$ corresponds to the $\mp$-equations \eqref{eq:btzhs1st}. In \eqref{eq:btzallpm} we have the relative polarization constants fixed by \eqref{eq:btzhs1st} to be
\begin{align}\label{eq:polhsbtz}
    C^{(\mp)}_{\omega l,(+)} \frac{\Gamma(a_{-\omega+is r_H,\mp l})}{\Gamma(a_{-\omega-is r_H,\mp l})}=(-)^s C^{(\mp)}_{\omega l,(-)} \frac{\Gamma(a_{\omega+is r_H,\pm l})}{\Gamma(a_{\omega-is r_H,\pm l})}\; .
\end{align}
Combining \eqref{eq:btzs0nh}, \eqref{eq:btzallpm} and \eqref{eq:polhsbtz}, we can deduce
\begin{align}\label{eq:btzhsnh}
    \left(\phi_{(+)}, \phi_{(-)} \right)^{(\mp)}_{\omega l}  (x\to -\infty) 
    \propto  B^{(\mp), \text{out}}_{\omega l} (0,(-)^s) \, e^{(-i \omega -sr_H)(t+x)}+B^{(\mp), \text{in}}_{\omega l}(1,0)\, e^{(-i \omega +sr_H)(t-x)}
\end{align}
where $(0,(-)^s) \, e^{(-i \omega -sr_H)(t+x)}$ and $(1,0)\, e^{(-i \omega +sr_H)(t-x)}$ are the outgoing and incoming waves respectively, with coefficients
\begin{align}
    B^{(\mp), \text{out}}_{\omega l}=\frac{\Gamma\left(\frac{i\omega}{r_H}+s \right)}{\Gamma\left(a_{-\omega-is r_H, \mp l} \right)\Gamma\left(a_{-\omega+i sr_H, \pm l} \right)}\; , \qquad  B^{(\mp), \text{in}}_{\omega l} =\frac{\Gamma\left(-\frac{i\omega}{r_H}+s \right)}{\Gamma\left(a_{\omega-is r_H, \pm l} \right)\Gamma\left(a_{\omega+i sr_H, \mp l} \right)} \; .
\end{align}
The ratio between the coefficients is again a pure phase and takes the form
\begin{align}\label{eq:btzhsSmatfull}
    \mathcal{S}^{(\mp)}_l(\omega) \equiv \frac{B^{(\mp), \text{out}}_{\omega l}}{B^{(\mp), \text{in}}_{\omega l}} = \mathcal{S}^{\text{BTZ}, (\mp )}_{s,l}(\omega) \, \mathcal{S}_{ l}^{\text{Rin}, ( s)} \left(\beta_H,\omega\right)
\end{align}
with $\mathcal{S}_{ l}^{\text{Rin}, ( s)} (\beta,\omega)$ defined in \eqref{appeq:rinhsSmat} and
\begin{align}\label{eq:btzhsSmat}
    \mathcal{S}^{\text{BTZ}, (\mp )}_{s,l}(\omega) =&\,  \mathcal{S}^{\text{BTZ}, (\mp ,L)}_{s,l}(\omega)\mathcal{S}^{\text{BTZ}, (\mp ,R)}_{s,l}(\omega) \;, \nn\\
    \mathcal{S}^{\text{BTZ}, (\mp ,L)}_{s,l}(\omega)\equiv \frac{\Gamma\left(a_{\omega \mp  i sr_H, l} \right)}{\Gamma\left(a_{-\omega\mp i s r_H, - l} \right)} \; , &\qquad \mathcal{S}^{\text{BTZ}, (\mp, R )}_{s,l}(\omega) \equiv \frac{\Gamma\left(a_{\omega \pm  i sr_H, -l} \right)}{\Gamma\left(a_{-\omega\pm i s r_H, l} \right)} \; ,
\end{align}
in accordance with the discussion at the end of Section \ref{sec:scattering}. QNM frequencies are the poles of the S-matrices $\mathcal{S}^{\text{BTZ}, (\mp )}_{s,l}(\omega)$, which can be summarized by
\begin{align}\label{eq:btzhspmqnm}
    z^{(\mp),L}_{nl}= l - 2\pi T_H i (\Delta\mp s+2n)  \qquad z^{(\mp),R}_{nl}=- l - 2\pi T_H i (\Delta \pm s+2n) \;,
\end{align}
while the anti-QNM frequencies are given by $\left(z^{(\mp),L/R}_{nl}\right)^*$, the zeros of $\mathcal{S}^{\text{BTZ}, (\mp )}_{s,l}(\omega)$. In a parity invariant theory where both $(\mp)$-QNMs \eqref{eq:btzhspmqnm} are included, the set of anti-QNMs is also spanned by $-z^{(\mp),L/R}_{nl}$, the zeros of $\mathcal{S}^{\text{BTZ}, (\pm )}_{s,l}(\omega)$.

\subsection{Euclidean continuation of the quasinormal modes}\label{sec:btzexam}

In this part we examine the Euclidean continuation of the QNMs. The analysis for anti-QNMs is analogous.

\subsubsection{Massive scalar}

As a warm-up illustration, we first look at the case of the massive scalar. At the QNM frequencies \eqref{eq:btzs0qnm}, with \eqref{eq:btzs0an}, \eqref{eq:btzscalarsol} we can write down the full explicit mode functions 
\begin{align}\label{eq:btzs0qnmfn}
    \phi^L_{n l}
    \propto & \, e^{-iz^L_{n,l} t} (\cosh\left(r_H x \right) )^{\frac{il}{ r_H}}\left(-\sinh \left(r_H x \right) \right)^{\Delta } \, _2F_1\left(-n ,\frac{iz^L_{nl}}{r_H}-n;\Delta ;-\sinh^2 \left(r_H x \right) \right) \;, \nn\\
    \phi^R_{n l}
    \propto & \, e^{-iz^R_{n,l} t} (\cosh\left(r_H x \right) )^{-\frac{il}{ r_H}}\left(-\sinh \left(r_H x \right) \right)^{\Delta } \, _2F_1\left(-n ,\frac{iz^R_{nl}}{r_H}-n;\Delta ;-\sinh^2 \left(r_H x \right) \right) \;.
\end{align}
We have suppressed the $e^{il\vartheta}$ dependence which is unimportant for the following. In deriving $\phi^R_{n l}(x)$ we have used
\begin{align}\label{eq:2f1iden1}
    \, _2F_1(a,b;c;z) &= (1-z)^{c-a-b}\, _2F_1(c-a,c-b;c;z) \;.
\end{align}
Following the DHS analyticity argument, we vary $m^2$ or $\Delta$ such that 
\begin{align}
    \frac{iz^L_{nl}}{r_H} = -|k|  \qquad \text{or} \qquad \frac{iz^R_{nl}}{r_H} = -|k|\; , \qquad k\in \mathbb{Z}\;.
\end{align}
At these (complex) mass values, upon Wick rotating $t\to-it_E = -i \frac{1}{r_H}\varphi$, the mode functions \eqref{eq:btzs0qnmfn} behave near the origin like
\begin{align}\label{eq:btzs0qnmorigin}
    \phi^L_{n l} (-it_E,x\to -\infty)&\propto e^{-i |k| \varphi} e^{(|k|+2 n)r_Hx} \, _2F_1\left(-n ,-|k|-n;\Delta ;-\frac{1}{4}e^{-2r_H x} \right) \nn\\
    \phi^R_{n l} (-it_E,x\to -\infty)&\propto e^{-i |k| \varphi} e^{(|k|+2 n)r_H x} \, _2F_1\left(-n ,-|k|-n;\Delta ;-\frac{1}{4}e^{-2r_H x} \right)\; .
\end{align}
To proceed, we will make use of another identity
\begin{align}\label{eq:2f1iden2}
    F(-m,b;c;z) &= \frac{\Gamma(b+m) \Gamma(c)}{\Gamma(b)\Gamma(c+m)}(-z)^{m} F\left(-m,1-c-m;1-b-m;\frac{1}{z} \right) \;,
\end{align}
where $m$ is a non-negative integer. It is important to note that when $b = -j$ for $j$ a non-negative integer, \eqref{eq:2f1iden2} holds if $m \le j$. Using this, one can show that for any $n=0,1,2,\dots $ and $k\in \mathbb{Z}$, \eqref{eq:btzs0qnmorigin} is equivalent to
\begin{align}
    \phi^L_{n l} (-it_E,x\to -\infty)\propto u^{|k|} \qquad \text{or} \qquad \phi^R_{n l} (-it_E,x\to -\infty)  \propto e^{|k|(r_H x-i\varphi)}= u^{|k|} \;.
\end{align}
We thus conclude that these modes obey the regularity condition \eqref{quan asym}.

\subsubsection{Massive higher spin}

Now, let us study the Wick-rotation of QNMs for a massive spin-$s$ field, for general $s\geq 1$. From \eqref{eq:btzhsnh}, we observe that at the QNM frequencies \eqref{eq:btzhspmqnm}, the incoming piece vanishes and we are left with the purely outgoing piece determined by $\phi_{(-)}$. Using \eqref{eq:btzallpm}, we first write down the explicit normal modes for $\phi_{(-)}$:
\begin{align}
	\phi_{\omega l (-)} \propto e^{ - s r_H t - i \omega t} \frac{(\cosh\left(r_H x \right) )^{\frac{il}{ r_H}}\left(-\sinh \left(r_H x \right) \right)^{\Delta }}{\left(-\tanh(r_H x)\right)^s} \, _2F_1\left(a_{\omega- isr_H, l} ,a_{-\omega + isr_H, l};\Delta ;-\sinh^2 \left(r_H x \right) \right) \;.
\end{align}
We have suppressed the $e^{il\vartheta}$ dependence which is unimportant for the following. The QNMs at frequencies \eqref{eq:btzhspmqnm} read explicitly
\begin{align}\label{eq:btzhsqnmfn}
     \phi^{(-),L}_{n l, (-)} &\propto \, e^{-i( z^{(-),L}_{nl} - isr_H )t} \frac{(\cosh\left(r_H x \right) )^{\frac{il}{ r_H}}\left(-\sinh \left(r_H x \right) \right)^{\Delta }}{\left(-\tanh(r_H x)\right)^s} \ {}_2F_{1}\left(-n,\frac{iz^{(-),L}_{nl}}{r_H}-n+ s; \Delta;-\sinh^2(r_H x) \right) \,, \nn\\
    \phi^{(-),R}_{n l, (-)} &\propto \, e^{-i( z^{(-),R}_{nl} - isr_H )t} \frac{(\cosh\left(r_H x \right) )^{-\frac{il}{ r_H}}\left(-\sinh \left(r_H x \right) \right)^{\Delta }}{\left(-\tanh(r_H x)\right)^s} \ {}_2F_{1}\left(-n-s,\frac{iz^{(-),R}_{nl}}{r_H}-n; \Delta;-\sinh^2(r_H x) \right) \;, \nn\\
    \phi^{(+),L}_{n l, (-)} &\propto \, e^{-i( z^{(+),L}_{nl} - isr_H )t} \frac{(\cosh\left(r_H x \right) )^{\frac{il}{ r_H}}\left(-\sinh \left(r_H x \right) \right)^{\Delta }}{\left(-\tanh(r_H x)\right)^s} \ {}_2F_{1}\left(-n-s,\frac{iz^{(+),L}_{nl}}{r_H}-n; \Delta;-\sinh^2(r_H x) \right) \,, \nn\\
    \phi^{(+),R}_{n l, (-)} &\propto \, e^{-i( z^{(+),R}_{nl} - isr_H )t} \frac{(\cosh\left(r_H x \right) )^{-\frac{il}{ r_H}}\left(-\sinh \left(r_H x \right) \right)^{\Delta }}{\left(-\tanh(r_H x)\right)^s} \ {}_2F_{1}\left(-n,\frac{iz^{(+),R}_{nl}}{r_H}-n+s; \Delta;-\sinh^2(r_H x) \right) \;,
\end{align}
Again, we vary $m^2$ or $\Delta$ such that 
\begin{align}\label{eq:btzhsvary}
    \frac{iz^{(\mp),L}_{nl}}{r_H} = -|k|  \qquad \text{or} \qquad \frac{iz^{(\mp),R}_{nl}}{r_H} = -|k|\; , \qquad k\in \mathbb{Z}\;.
\end{align}
At these (complex) values of masses, upon Wick rotating $t\to-it_E = -i \frac{1}{r_H}\varphi$, the mode functions \eqref{eq:btzhsqnmfn} behave near the origin like
\begin{align}\label{eq:Eucbtzmode}
    \phi^{(-),L}_{n l,(u)} (-it_E,x\to -\infty)&\propto e^{-i (|k|- s) \varphi} e^{(|k|+2 n- s)r_H x} \, _2F_1\left(-n ,-|k|-n+ s;\Delta ;-\frac{1}{4}e^{-2r_H x} \right) \nn\\
    \phi^{(-),R}_{n l,(u)} (-it_E,x\to -\infty)&\propto e^{-i (|k|- s) \varphi} e^{(|k|+2 n- s)r_H x} \, _2F_1\left(-n-s ,-|k|-n;\Delta ;-\frac{1}{4}e^{-2r_H x} \right)\; ,\nn\\
    \phi^{(+),L}_{n l,(u)} (-it_E,x\to -\infty)&\propto e^{-i (|k|- s) \varphi} e^{(|k|+2 n- s)r_H x} \, _2F_1\left(-n-s ,-|k|-n;\Delta ;-\frac{1}{4}e^{-2r_H x} \right) \nn\\
    \phi^{(+),R}_{n l,(u)} (-it_E,x\to -\infty)&\propto e^{-i (|k|- s) \varphi} e^{(|k|+2 n- s)r_H x} \, _2F_1\left(-n ,-|k|-n+s;\Delta ;-\frac{1}{4}e^{-2r_H x} \right)\; .
\end{align}
Here we recall that under the Wick rotation, the $-$-component becomes the $u$-component.
Using \eqref{eq:2f1iden2}, one can show that for any $|k|\geq s$, all four mode functions \eqref{eq:Eucbtzmode} are regular near the origin
\begin{align}
    \phi^{(\mp),L}_{n l, (u)} (-it_E,x\to -\infty) \propto \phi^{(\mp),R}_{n l, (u)} (-it_E,x\to -\infty) \propto e^{(|k|-s)r_H x}e^{-i (|k|- s) \varphi} \propto u^{|k|-s} \;. 
\end{align}
The case for $|k|<s$ is more intricate. On the one hand, we always have
\begin{align}
    \phi^{(+),L}_{n l, (u)} (-it_E,x\to -\infty) \propto \phi^{(-),R}_{n l, (u)} (-it_E,x\to -\infty)\propto e^{(s-|k|)r_H x}e^{-i (|k|- s) \varphi} \propto {\bar u}^{s-|k|} \; ,
\end{align}
so that these are regular at the origin. On the other, when $|k|+n \geq s >|k|$, 
\begin{align}\label{eq:btzhsks}
    \phi^{(-),L}_{n l, (u)} (-it_E,x\to -\infty) \propto \phi^{(+),R}_{n l, (u)} (-it_E,x\to -\infty) \propto e^{(s-|k|)r_H x}e^{-i (|k|- s) \varphi}
    =
    {\bar u}^{s-|k|} 
    \; ,
\end{align}
while for $|k|+n<s$,
\begin{align}\label{eq:btzhsk0}
    \phi^{(-),L}_{n l, (u)} (-it_E,x\to -\infty) \propto \phi^{(+),R}_{n l, (u)} (-it_E,x\to -\infty) \propto e^{||k|-s|r_H x}e^{-i (|k|- s) \varphi}
    =
    u^{-(s-|k|)} \; .
\end{align}
We can see that \eqref{eq:btzhsks} is a regular behavior while \eqref{eq:btzhsk0} is not.

To summarize, for a fixed $k\in \mathbb{Z}$, any QNM can Wick-rotate to a regular Euclidean mode with $U(1)$ quantum number $|k|$ at complex masses \eqref{eq:btzhsvary}, except for those with frequencies 
\begin{align}\label{eq:btzedgefre}
    z^{(-),L}_{nl}= l - 2\pi T_H i (\Delta- s+2n)  \qquad z^{(+),R}_{nl}=- l - 2\pi T_H i (\Delta - s+2n) \;, \qquad n< s-|k| \; .
\end{align}
The irregularity of such modes agrees with the case of $s=2$ first pointed out in \cite{Castro:2017mfj}.


\subsection{Euclidean path integral}\label{sec:edgebtzpi}

\subsubsection{Quasinormal mode character and renormalized bulk partition function}



With the QNM spectrum \eqref{eq:btzhspmqnm} we can compute the QNM character 
\begin{align}\label{btz char}
	\chi_{[\Delta,s]}^\text{BTZ} (t) =\sum_{l\in \mathbb{Z}} \sum_{n=0}^\infty \sum_{\pm}\left( e^{-i z^L_{n,l,\pm}t} +e^{-i z^R_{n,l,\pm} t}\right)=  \frac{4\pi  e^{-2\pi T_H\Delta t}}{1-e^{-4\pi T_H t}}2\cosh (2\pi T_H s t)\sum_{k\in\mathbb{Z}}  \delta(t-2\pi k)\; .
\end{align}
Substituting this into \eqref{eq:DHS} yields the renormalized bulk partition function
\begin{align}\label{eq:HS BTZ pf}
	\log \widetilde{Z}^\text{BTZ}_\text{bulk} \left[\Delta,s\right]
	=&\sum_{k=1}^\infty \frac{q_k^s+q_k^{-s}}{k}\frac{q_{k}^{\Delta}}{(1-q_{k})^2} \; ,\qquad q_{k}=e^{-(2\pi)^2 T_H k}\; .
\end{align}
Note that this is not related to the $TAdS_3$ expression \eqref{appeq:TAdS} through the modular transformation \eqref{eq:mod trans} for any $s\geq 1$.

\subsubsection{Edge partition function}

As we checked explicitly in Section \ref{sec:btzexam}, for a fixed $k\in \mathbb{Z}$, QNMs with frequencies \eqref{eq:btzedgefre} cannot Wick-rotate to a regular Euclidean mode with $U(1)$ quantum number $|k|$ at complex masses \eqref{eq:btzhsvary}. They contribute to the integrand \eqref{eq:generaledge} as
\begin{align}
    &\frac{1}{2t}\sum_{k=-(s-1)}^{s-1} \sum_{l\in\mathbb{Z}}\sum_{n=0}^{s-1-|k|} \, \left( e^{-\left(2\pi T_H |k|+i z^{(-),L}_{nl}\right)t}+e^{-\left(2\pi T_H |k|+i z^{(+),R}_{nl}\right)t} \right)\nn\\
   =&\frac{2\pi e^{-2 \pi  T_H
   \Delta t}}{t} \sum_{j\in \mathbb{Z}}\delta(t-2\pi j) \frac{ e^{-2 \pi  s T_H
   t}+e^{2 \pi  s T_H
   t}-2}{\left(1-e^{-2 \pi  t T_H}\right){}^2} 
\end{align}
Substituting this into \eqref{eq:generaledge} yields the BTZ edge partition function
\begin{equation}\label{eq:btzedge}
    \log Z_{\text{edge}}= \sum_j \frac{q^s_j + q^{-s}_j-2}{j}\frac{q^\Delta_j}{(1-q_j)^2}\;.
\end{equation}

\subsubsection{The full Euclidean path integral}

Taking the difference between the renormalized bulk partition function \eqref{eq:HS BTZ pf} and the edge partition function \eqref{eq:btzedge}, we obtain the full path integral 
\begin{equation}\label{eq:HS BTZ PI}
    \log Z^\text{BTZ}_\text{PI}= \log \widetilde{Z}^\text{BTZ}_\text{bulk}-\log Z^\text{BTZ}_\text{edge} = \sum^\infty_{k=1} \frac{2}{k} \frac{q^\Delta_k}{(1-q_k)^2},
\end{equation}
which precisely equals the $TAdS_3$ result \eqref{appeq:TAdS} after the modular transformation \eqref{eq:mod trans}. In \cite{datta_higher_2012} the authors found a prescription to modify the DHS formula so that the result \eqref{eq:HS BTZ PI} was reproduced. Our discussion in Section \ref{sec:dhsspin} gave a justification for their prescription.

\section{Example: Massive vector on Nariai}\label{sec:nariai}

In this section we study a free massive vector $A_\mu$ on Nariai spacetime ($d\geq 3$): 
\begin{align}\label{eq:narcoords}
	ds^2=-\left(1-y^2 \right)dt^2 +\frac{\ell^2_N }{1-y^2}\;dy^2+r_N^2\, d\Omega_{d-1}^2\; , \quad -1<y<1\; .
\end{align}
Here $\ell_N$ and $r_N$ are related to the dS length $\ell_\text{dS}$ through
\begin{align}\label{eq:nariaipara}
 \ell_N\equiv \frac{\ell_\text{dS}}{\sqrt{d}} \;, \qquad r_N \equiv  \sqrt{\frac{d-2}{d}}\ell_\text{dS} \;, \qquad \ell_\text{dS} \equiv \sqrt{\frac{d(d-1)}{2\Lambda}}  \; .
\end{align}
This geometry is locally $dS_2\times S^{d-1}$, with isometry group $SO(1,2)\times SO(d)$. There are two horizons (cosmological and black hole) at $y =\pm 1$ with the same Hawking temperatures $T_N=\frac{1}{2\pi \ell_N}$. Note that this temperature is higher than the temperature $T_\text{dS}=\frac{1}{2\pi \ell_\text{dS}}$ for pure de Sitter. 


Upon Wick-rotating $t\to -i t_E$ and identifying $t_E \sim t_E +2\pi \ell_N$, the geometry \eqref{eq:narcoords} becomes $S^2\times S^{d-1}$. The 1-loop path integral for a massive vector on such a geometry is
\begin{align}
    Z_\text{PI} = \int \mathcal{D}A \, e^{-S[A]} \quad , \qquad
S[A]=\int_{S^2 \times S^{d-1}} \left(\frac{1}{4}F_{\mu\nu}F^{\mu\nu} +\frac{m^2}{2} A_\mu A^\mu\right) \, .
\end{align}
The derivation in \cite{Law:2020cpj} is readily carried over to this case: 
\begin{align}\label{eq:nariai vec det}
Z_\text{PI}=Z^T_\text{PI}\, Z^L_\text{PI} 
\end{align}
with
\begin{align}
    Z^T_\text{PI} = \det(-\nabla_{(1)}^2 + m^2 +\frac{1}{\ell_N^2})^{-1/2} \; , \qquad Z^L_\text{PI} =(m^2)^{1/2} \; .
\end{align}
Here $-\nabla_{(1)}^2$ is the Laplacian acting on transverse vector fields on $S^2\times S^{d-1}$. We have used the fact that the $S^2$ and $S^{d-1}$ factors in the Euclidean Nariai geometry have respective radii $\ell_N$ and $r_N$ defined in \eqref{eq:nariaipara}. The factor $Z^L_\text{PI}$ comes from the integration over the off-shell longitudinal modes and corresponds to the normalizable constant scalar mode \cite{Anninos:2020hfj,Law:2020cpj}. Our general discussions in Section \ref{sec:dhsspin} apply to $Z^T_\text{PI}$. We will comment on $Z^L_\text{PI}$ when we directly compute $Z_\text{PI}$ in Section \ref{app:nareucl}.

\subsection{Explicit solutions and quasinormal modes}\label{app:narqnm}


The Proca equation of motion $\nabla^\mu F_{\mu\nu}=m^2 A_\nu$ on \eqref{eq:narcoords} is equivalent to 
\begin{align}\label{eq:nareom}
    \left(-\frac{1}{\ell_N^2}\nabla_{dS_2}^2-\frac{1}{r_N^2}\nabla^2_{S^{d-1}}+m^2 +  \frac{1}{\ell_N^2}\right)A_\mu &=0 \; .
\end{align}
In deriving this, it is important to use the relation \eqref{eq:nariaipara} between $\ell_N$ and $r_N$. In these expressions, $\nabla^2_{dS_2}$ and $\nabla^2_{S^{d-1}}$ are the Laplacians on the {\it unit} $dS_2$ and $S^{d-1}$ respectively; the former acts on $(A_t,A_y)$ as a vector and $A_i$ as a scalar, while the latter acts on $(A_t,A_y)$ as scalars and $A_i$ as a vector. All components are related through the transversality condition
\begin{equation}\label{eq:nartrans}
    \nabla^\mu A_\mu=-\frac{1}{1-y^2}\partial_t A_t+\frac{1-y^2}{\ell_N^2}\partial_y A_y +\frac{1}{r_N^2}\nabla^i_{S^{d-1}} A_i=0\; .
\end{equation}
There are three types of solutions according to their $SO(d)$ transformation properties. In Appendix \ref{app:harmonics}, we collect some basic facts about scalar and vector spherical harmonics that are useful for our analysis.

\paragraph{Vector mode}

This tower of solutions take the form 
\begin{align}\label{eq:vecans}
    A_t=A_y=0 \;, \qquad A_i  = e^{-i\omega t} R^V(y) Y^{d-1}_{l, i}(\Omega) \; , \qquad l\geq 1\; , 
\end{align}
where $Y^{d-1}_{l, i}(\Omega)$ are vector spherical harmonics on $S^{d-1}$. The transversality condition as well as the $y,t$-components of the equation of motion \eqref{eq:nareom} are automatically satisfied. The $i$-component of \eqref{eq:nareom} implies that
\begin{equation}\label{eq:narReq}
     (1-y^2)\partial^2_y R -2y\partial_y R+\left(\frac{\omega^2 \ell^2_N}{1-y^2}-\Delta_{V,l} \bar{\Delta}_{V,l}\right)R=0\; ,
\end{equation}
where 
\begin{equation}
   \Delta_{V,l} = \frac12 + i\, \nu_{V,l} \;,\qquad \nu_{V,l} =\sqrt{ \ell^2_N  m_{V,l}^2+\frac34} \;, \qquad m_{V,l}^2\equiv m^2 + \frac{l(l+d-2)-1}{r_N^2}\;,
\end{equation}
and $\bar \Delta_{V,l} \equiv 1-\Delta_{V,l} $. There are two linearly independent solutions to \eqref{eq:narReq}: 
\begin{equation}\label{eq:narRsol}
  R^{V,\text{even}}_{\omega l} = (1-y^2)^{-\frac{i\omega \ell_N}{2}}\; {}_2F_1\left(\frac{\Delta_{V,l}-i\omega \ell_N}{2},\frac{\bar{\Delta}_{V,l}-i\omega \ell_N}{2};\frac12,y^2\right)
\end{equation}
and
\begin{equation}
\begin{split}\label{eq:narRsol2}
  R^{V,\text{odd}}_{\omega l} = y\; (1-y^2)^{-\frac{i\omega \ell_N}{2}}\;{}_2F_1\left(\frac{1+\Delta_{V,l}-i\omega \ell_N}{2},\frac{1+\bar{\Delta}_{V,l}-i\omega \ell_N}{2};\frac32,y^2\right) \; .
  \end{split}
\end{equation}
Both modes are regular at the location of the observer $y=0$ and thus should be included as solutions. The label even/odd denotes the parity under $y\to -y$. Expanding near the horizons in terms of the tortoise coordinate $x = \frac{\ell_N}{2}\log\frac{1+y}{1-y}$ we find respectively
\begin{equation}\label{eq:narvecnh1}
    R^{V,\text{even}}(|x|\to \infty) \propto\; \frac{\Gamma(i\omega \ell_N)}{\Gamma\left(\frac{\Delta_{V,l}+i\omega \ell_N}{2}\right)\Gamma\left(\frac{\bar{\Delta}_{V,l}+i\omega\ell_N}{2}\right)} e^{i\omega |x|}+ \left( \omega \to -\omega\right)\; ,
\end{equation}
and
\begin{equation}\label{eq:narvecnh2}
R^{V,\text{odd}}(|x|\to \infty) \propto \; \frac{\Gamma(i\omega \ell_N)}{\Gamma\left(\frac{1+\Delta_{V,l}+i\omega \ell_N}{2}\right)\Gamma\left(\frac{1+\bar{\Delta}_{V,l}+i\omega\ell_N}{2}\right)} e^{i\omega |x|}+ \left( \omega \to -\omega\right)\;.
\end{equation}
Therefore, at QNM frequencies
\begin{align}\label{eq: scal1qnmV}
    i z^V_{\Delta n l} \ell_N = \Delta_{V,l} + n\; , \qquad i z^V_{\bar\Delta n l} \ell_N = \bar{\Delta}_{V,l} + n\; , \qquad n=0,1,2,\dots \; ,
\end{align}
the even (odd) modes \eqref{eq:narvecnh1} are purely outgoing at both horizons when $n$ is even (odd). At these frequencies, one can solve \eqref{eq:narReq} to get
\begin{align}\label{eq:narvecqnm}
	R^{V}_{\Delta n l}(y)= P_{-\Delta_{V,l}}^{n+\Delta_{V,l}}(y)\quad \text{and} \quad R^{V}_{\bar\Delta n l}(y)= P_{-\bar\Delta_{V,l}}^{n+\bar\Delta_{V,l}}(y)
\end{align}
where $n=0, 1, 2, \dots$. Alternatively, \eqref{eq:narvecqnm} can obtained (up to an overall normalization) by substituting \eqref{eq:narvecqnm} into \eqref{eq:narRsol} (\eqref{eq:narRsol2}) when $n$ is even (odd) together with the relation between between hypergeometric and associated Legendre functions (See for instance \cite[$\S$14.3(iii)]{NIST:DLMF}). The spectrum of anti-QNM can be solved in a similar way, and the explicit mode functions can be obtained by flipping $t\to -t$ in the QNM ones.

\paragraph{Scalar mode I}

This tower of scalar solutions takes the form
\begin{equation}\label{eq:narscalarI}
    A_t = i \frac{1-y^2}{\ell_N} R'(y) e^{-i\omega t}Y^{d-1}_{l}(\Omega)\; , \quad A_y = \frac{\omega \ell_N}{1-y^2} R(y) e^{-i\omega t} Y^{d-1}_{l}(\Omega)\; , \quad A_i = 0 \; , \quad l\geq 0,
\end{equation}
where $Y^{d-1}_{l}(\Omega)$ are scalar spherical harmonics on $S^{d-1}$. Plugging this ansatz into \eqref{eq:nareom} and \eqref{eq:nartrans}, one finds that $R$ satisfies \eqref{eq:narReq}, with $\Delta_{V,l}$ replaced by $\Delta_{S,l}$ defined as
\begin{equation}\label{eq:scaDel}
    \Delta_{S,l} = \frac12 + i\nu_{S,l}\; ,\qquad \nu_{S,l} = \sqrt{m^2_{S,l} \ell^2_N-\frac14} =\sqrt{m^2\ell^2_N+l(l+d-2)-\frac14}\;, \qquad \bar \Delta_l \equiv 1- \Delta_l \; .
\end{equation}
We essentially get a KK-tower of $dS_2$ vectors with masses $m^2_{S,l}$. The subsequent analysis is the same as before but with $\Delta_{V,l}$ replaced by $\Delta_{S,l}$. In particular, we can immediately write down the QNM frequencies 
\begin{equation}\label{eq: scal1qnms}
    i z^{(S,1)}_{\Delta n l} \ell_N = \Delta_{S,l} + n\; , \qquad i z^{(S,1)}_{\bar\Delta n l} \ell_N = \bar{\Delta}_{S,l} + n\; , \qquad n=0,1,2,\dots \;,
\end{equation}
and the mode functions 
\begin{align}\label{eq:narqnmS1}
 	R^{(S,1)}_{\Delta n l}(y)= P_{-\Delta_{S,l}}^{n+\Delta_{S,l}}(y)\qquad \text{and} \qquad R^{(S,1)}_{\bar\Delta n l}(y)= P_{-\bar\Delta_{S,l}}^{n+\bar\Delta_{S,l}}(y) \;.
\end{align}
The mode functions of the anti-QNMs can be obtained by flipping $t\to -t$ in the QNM ones.




\paragraph{Scalar mode II}

Another tower of scalar solutions are obtained with the ansatz
\begin{equation}\begin{split}
   A_t &= -i \omega R(y)\; e^{-i\omega t}\; Y^{d-1}_{l}(\Omega)\;,\qquad A_y =  R'(y) \;e^{-i\omega t}\;Y^{d-1}_{l}(\Omega)\;,\\
   &\qquad A_i = \mathcal{C}_{l} R(y) \;e^{-i\omega t}\; \frac{\partial_i Y^{d-1}_{l}(\Omega)}{l(l+d-2)}\;,\qquad \qquad l\geq1\; .
   \end{split}
\end{equation}
The equations of motion again are satisfied when $R$ solves \eqref{eq:narReq} with $\Delta_{V,l}$ replaced by $\Delta_{S,l}$ defined in \eqref{eq:scaDel}. For this set of solutions,  the transversality constraint \eqref{eq:nartrans} implies that the respective divergences along the $dS_2$ and $S^{d-1}$ directions cancel each other, with the polarization constant $\mathcal{C}_{l}$ fixed to be
\begin{equation}
   \Delta_{S,l} \bar{\Delta}_{S,l} = \frac{\mathcal{C}_{l}}{d-2}\; .
\end{equation}
Again the QNM frequencies are given by replacing $\Delta_{V,l}\to\Delta_{S,l}$ in \eqref{eq: scal1qnmV}:
\begin{equation}\label{eq: scal2qnms}
    i z^{(S,2)}_{\Delta n l} \ell_N = \Delta_{S,l} + n, \quad i z^{(S,2)}_{\bar\Delta n l} \ell_N = \bar{\Delta}_{S,l} + n, \quad n=0,1,2,\dots \;.
\end{equation}
These are identical to \eqref{eq: scal1qnms} but we stress that here $l\geq 1$. The mode functions are 
\begin{align}\label{eq:narqnmS2}
	R^{(S,2)}_{\Delta n l}(y)= P_{-\Delta_{S,l}}^{n+\Delta_{S,l}}(y)\quad \text{and} \quad R^{(S,2)}_{\bar\Delta n l}(y)= P_{-\bar\Delta_{S,l}}^{n+\bar\Delta_{S,l}}(y) \;. 
\end{align}
Again, the anti-QNM functions can be obtained by flipping $t\to -t$ in the QNM ones.

\subsubsection{Renormalized bulk partition function}

With the QNM spectra \eqref{eq: scal1qnmV}, \eqref{eq: scal1qnms} and \eqref{eq: scal2qnms}, we can immediately write down the QNM character 
\begin{equation}\label{eq:narbulk}
    \chi(t) =\sum_{l\geq 1}\left( D^d_{l,1}\frac{q^{\Delta_{V,l}} + q^{\bar{\Delta}_{V,l}}}{1-q} + D^d_l\frac{q^{\Delta_{S,l}} + q^{\bar{\Delta}_{S,l}}}{1-q}\right)+\sum_{l\geq0}D^d_l\frac{q^{\Delta_{S,l}} + q^{\bar{\Delta}_{S,l}}}{1-q} \; .
\end{equation}
Here we have defined $q\equiv e^{-t/\ell_N}$. Plugging this into \eqref{eq:DHS} then gives
\begin{align}\label{eq:narrenbulk}
    \log \widetilde{Z}^T_{\text{bulk}}= \int_0^\infty& \frac{dt}{2t}\frac{1+q}{1-q}\left[\sum_{l\geq 1}\left( D^d_{l,1}\frac{q^{\Delta_{V,l}} + q^{\bar{\Delta}_{V,l}}}{1-q} + D^d_l\frac{q^{\Delta_{S,l}} + q^{\bar{\Delta}_{S,l}}}{1-q}\right)  +\sum_{l\geq0}D^d_l\frac{q^{\Delta_{S,l}} + q^{\bar{\Delta}_{S,l}}}{1-q}\right] \;. 
\end{align}


\subsection{Quasinormal modes and eigenfunctions on Euclidean Nariai}\label{appsub:narcheck}

\subsubsection{Spectrum for the vector Laplacian on $S^2\times S^{d-1}$}\label{sec:spec}

The eigenfunctions of the Laplace operator $-\nabla_{(1)}^2$ in \eqref{eq:nariai vec det} can be easily obtained by combining together the spherical harmonics on $S^2$ and $S^{d-1}$. We summarize this below, where we use $i,j,k$ and $I,J,K$ to denote indices on the $S^{d-1}$ and $S^2$ factors respectively.

\paragraph{Vector type} 

These eigenfunctions take the form
\begin{align}\label{eq:NeigfnV}
    A_\varphi=A_\theta=0 \;, \qquad A_i  = Y^2_{Lp} (\theta,\varphi) \,Y^{d-1}_{l,i}(\Omega)  \; , \qquad  L \geq 0\; ,\qquad  -L\leq p\leq L \; ,\qquad  l\geq 1
\end{align}
with eigenvalues
\begin{align}\label{eq:NeigvalV}
    \lambda^{(V)}_{L,l} = \frac{L(L+1)}{\ell_N^2}+\frac{l(l+d-2)-1}{r_N^2} 
\end{align}
and degeneracy $D_L^3 D_{l,1}^d$.

\paragraph{Scalar type I}

The first type of scalar eigenfunctions take the form
\begin{align}\label{eq:NeigfnS1}
    \left(A_\varphi,A_\theta\right)=Y^2_{Lp,I}(\theta,\varphi)\,Y^{d-1}_{l}(\Omega) \; , \qquad A_i =0\;, \qquad L \geq 1\; , \qquad -L\leq p\leq L\; , \qquad l\geq 0\;,
\end{align}
with eigenvalues
\begin{align}\label{eq:NeigvalS1}
    \lambda^{(S,1)}_{L,l} = \frac{L(L+1)-1}{\ell_N^2}+\frac{l(l+d-2)}{r_N^2}
\end{align}
and degeneracy $D_L^3 D_{l}^d$.

\paragraph{Scalar type II}

Another type of scalar eigenfunctions take the form
\begin{align}\label{eq:NeigfnS2}
\left(A_\varphi,A_\theta\right)=&\partial_I Y^2_{Lp}(\theta,\varphi)\,Y^{d-1}_{l}(\Omega) \; , \qquad A_i = C_{Ll} \, Y^2_{Lp} (\theta,\varphi)\, \partial_i Y^{d-1}_{l}(\Omega)\;, \nn\\
&\qquad L \geq 1\; , \qquad -L\leq p\leq L\; , \qquad l\geq 1\;,
\end{align}
with  eigenvalues 
\begin{align}\label{eq:NeigvalS2}
    \lambda^{(S,2)}_{L,l} = \frac{L(L+1)-1}{\ell_N^2}+\frac{l(l+d-2)}{r_N^2}= \frac{L(L+1)}{\ell_N^2}+\frac{l(l+d-2)-(d-2)}{r_N^2}
\end{align}
and degeneracy $D_L^3 \, D_{l}^d$. Here $C_{Ll}$ is a relative constant fixed by the Casimir equation.

\subsubsection{Euclidean continuation of the quasinormal modes}

Since we have the exact expressions for the QNMs discussed in Section \ref{app:narqnm} and the Euclidean eigenfunctions summarized above, we can directly compare them after the Wick rotation $t\to-it_E$.  We will ignore overall normalization constants unimportant for this analysis. Also, for a better comparison we change to the variables $y=\cos \theta$ and $\varphi = \frac{1}{\ell_N}t_E$.

\paragraph{Vector type}

For every $l\geq 1$, whenever the mass $m^2$ is varied such that
\begin{align}
    i z^V_{\Delta n l} \ell_N =n+\Delta_{V,l}=-|k| \qquad \text{or} \qquad i z^V_{\bar\Delta n l} \ell_N =n+\bar\Delta_{V,l}=-|k| \; , \qquad k\in\mathbb{Z} \; ,
\end{align}
upon Wick rotation $t\to -i t_E$, the QNM with frequency \eqref{eq: scal1qnmV} becomes
\begin{align}\label{eq:wickqnm}
    A_\varphi=A_\theta=0 \;, \qquad A_i  = Y^2_{n+|k|,-|k|} (\theta,\varphi) \,Y^{d-1}_{l,i}(\Omega) 
\end{align}
while the anti-QNM given by flipping $t\to -t$ becomes
\begin{align}\label{eq:wickaqnm}
    A_\varphi=A_\theta=0 \;, \qquad A_i  = Y^2_{n+|k|,|k|} (\theta,\varphi) \,Y^{d-1}_{l,i}(\Omega)  \;.
\end{align}
It is clear that running over $n=0,1,2,\dots$ and $k\in\mathbb{Z}$, \eqref{eq:wickqnm} and \eqref{eq:wickaqnm} span the set of eigenfunctions of the vector type \eqref{eq:NeigfnV}.

\paragraph{Scalar type I}

For every $l\geq 0$, whenever the mass $m^2$ is varied such that
\begin{align}
    i z^{(S,1)}_{\Delta n l} \ell_N =n+\Delta_{S,l}=-|k| \qquad \text{or} \qquad i z^{(S,1)}_{\bar\Delta n l} \ell_N =n+\bar\Delta_{S,l}=-|k| \; , \qquad k\in\mathbb{Z} \; ,
\end{align}
the QNM with frequency \eqref{eq: scal1qnms} upon Wick rotation $t\to -i t_E$ becomes
\begin{align}\label{eq:wickqnms}
    \left(A_\varphi,A_\theta\right)=Y^2_{n+|k|,-|k|,I}(\theta,\varphi)\,Y^{d-1}_{l}(\Omega) \; , \qquad A_i =0\;,
\end{align}
while the anti-QNM given by flipping $t\to -t$ becomes
\begin{align}\label{eq:wickaqnms}
    \left(A_\varphi,A_\theta\right)=Y^2_{n+|k|,|k|,I}(\theta,\varphi)\,Y^{d-1}_{l}(\Omega) \; , \qquad A_i =0 \;.
\end{align}
Running over $n=1,2,\dots$ and $k\in\mathbb{Z}$, \eqref{eq:wickqnm} and \eqref{eq:wickaqnm} span the set of eigenfunctions of the scalar type I \eqref{eq:NeigfnS1}. Notice that when $n=k=0$, the Wick-rotated mode does not belong to this set.

\paragraph{Scalar type II}

For every $l\geq 1$, whenever the mass $m^2$ is varied such that
\begin{align}
    i z^{(S,2)}_{\Delta n l} \ell_N =n+\Delta_{S,l}=-|k| \qquad \text{or} \qquad i z^{(S,2)}_{\bar\Delta n l} \ell_N =n+\bar\Delta_{S,l}=-|k| \; , \qquad k\in\mathbb{Z} \; ,
\end{align}
the QNM with frequency \eqref{eq: scal2qnms} upon Wick rotation $t\to -i t_E$ becomes
\begin{align}\label{eq:wickqnms2}
    \left(A_\varphi,A_\theta\right)=\partial_I Y^2_{n+|k|,-|k|}(\theta,\varphi)\,Y^{d-1}_{l}(\Omega) \; , \qquad A_i = C_{n+|k|,l} \, Y^2_{n+|k|,-|k|} (\theta,\varphi)\, \partial_i Y^{d-1}_{l}(\Omega)\;,
\end{align}
while the anti-QNM given by flipping $t\to -t$ becomes
\begin{align}\label{eq:wickaqnms2}
    \left(A_\varphi,A_\theta\right)=\partial_I Y^2_{n+|k|,|k|}(\theta,\varphi)\,Y^{d-1}_{l}(\Omega) \; , \qquad A_i = C_{n+|k|,l} \, Y^2_{n+|k|,|k|} (\theta,\varphi)\, \partial_i Y^{d-1}_{l}(\Omega)\;,
\end{align}
In these expressions, $C_{Ll}$ is as defined in \eqref{eq:NeigfnS2}. Running over $n=1,2,\dots$ and $k\in\mathbb{Z}$, \eqref{eq:wickqnms2} and \eqref{eq:wickaqnms2} span the set of eigenfunctions of the scalar type II \eqref{eq:NeigfnS2}. Notice that when $n=k=0$, the Wick-rotated mode does not belong to this set.

\subsubsection{Edge partition function}

From the last section, we see that all QNMs Wick-rotate to Euclidean modes for the correct value of masses, except the $n=0$ modes of both scalar type I and II with frequencies
\begin{equation}
    i z^{(S,1)}_{\Delta n l} \ell_N = \Delta_{S,l} \; , \qquad i z^{(S,1)}_{\bar\Delta n l} \ell_N = \bar{\Delta}_{S,l}  \;, \qquad l\geq 0 \; ,
\end{equation}
and 
\begin{equation}
    i z^{(S,2)}_{\Delta, n=0, l} \ell_N = \Delta_{S,l} \; , \qquad i z^{(S,2)}_{\bar\Delta, n=0, l} \ell_N = \bar{\Delta}_{S,l}  \;, \qquad l\geq 1 
\end{equation}
respectively. As we saw these modes do not Wick-rotate to the $k=0$ Euclidean modes for any value of masses, and contribute to the edge partition function as
\begin{align}\label{eq:narrenedge}
    \log Z^{T}_{\text{edge}}=&\int_0^\infty \frac{dt}{2t} \left(\sum_{l\geq 1}+\sum_{l\geq0}\right)D^d_l\left(q^{\Delta_{S,l}} + q^{\bar\Delta_{S,l}}\right)  \; .
\end{align}

\subsection{Euclidean path integral}\label{app:nareucl}

With the eigenvalues and degeneracies of the spin-1 Laplacian on $S^2\times S^{d-1}$, we write down the 1-loop path integral
\begin{align}\label{eq:narloopstart}
    \log Z_\text{PI}
     = &\int_0^\infty \frac{d\tau}{2\tau} e^{-\frac{\epsilon^2}{4\tau}}\Bigg[\sum_{l\geq1} \sum_{L\geq0} D^d_{l,1} D_L^3 \, e^{- \left(\lambda^{(V)}_{L,l}+m^2+\frac{1}{\ell_N^2}\right) \tau } \nn\\
      & \qquad + \left(\sum_{l\geq1}  + \sum_{l\geq 0}  \right)\sum_{L\geq1} D^d_{l} D_L^3 e^{-\left(\lambda^{(S)}_{L,l} +m^2+\frac{1}{\ell_N^2}\right) \tau  }-e^{-m^2\tau }\Bigg] \; ,
\end{align}
where we have abbreviated $\lambda^{(S,1)}_{L,l} =\lambda^{(S,2)}_{L,l} =\lambda^{(S)}_{L,l} $. The last term in the bracket comes from the factor $Z^L_{\text{PI}}=(m^2)^{1/2}$ in \eqref{eq:nariai vec det}. Here we have inserted the UV regulator $e^{-\frac{\epsilon^2}{4\tau}}$ so that this integral is convergent for $\epsilon>0$.

\begin{figure}[H]
    \centering
   \begin{subfigure}{0.3\textwidth}
            \centering
            \includegraphics[width=\textwidth]{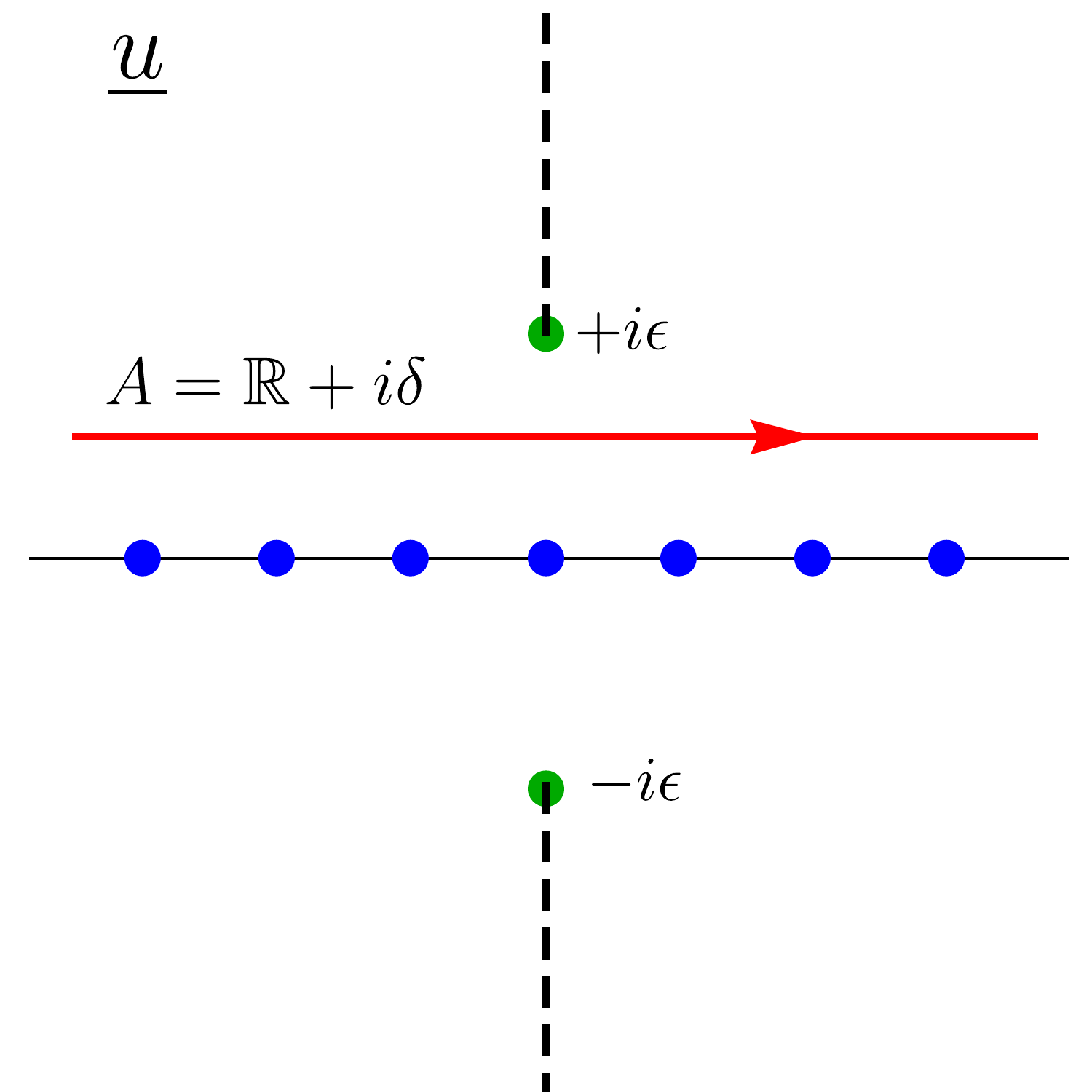}
            \caption[]%
            {{\small original contour}
             }    
        \end{subfigure}
        \begin{subfigure}{0.3\textwidth}  
            \centering 
            \includegraphics[width=\textwidth]{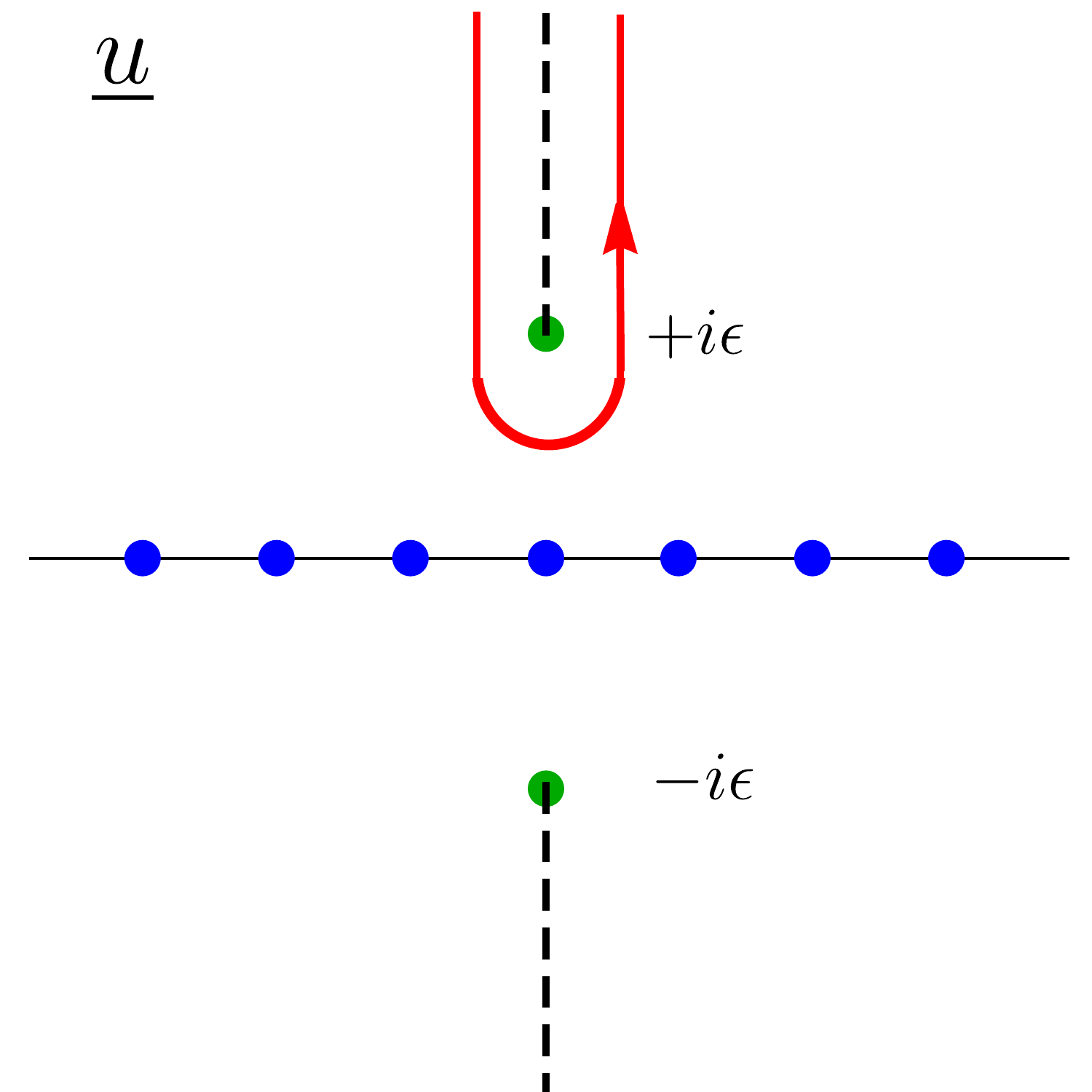}
            \caption[]%
            {{\small \centering  folded contour}}   
        \end{subfigure}
        \begin{subfigure}{0.3\textwidth}  
            \centering 
            \includegraphics[angle=270,width=\textwidth]{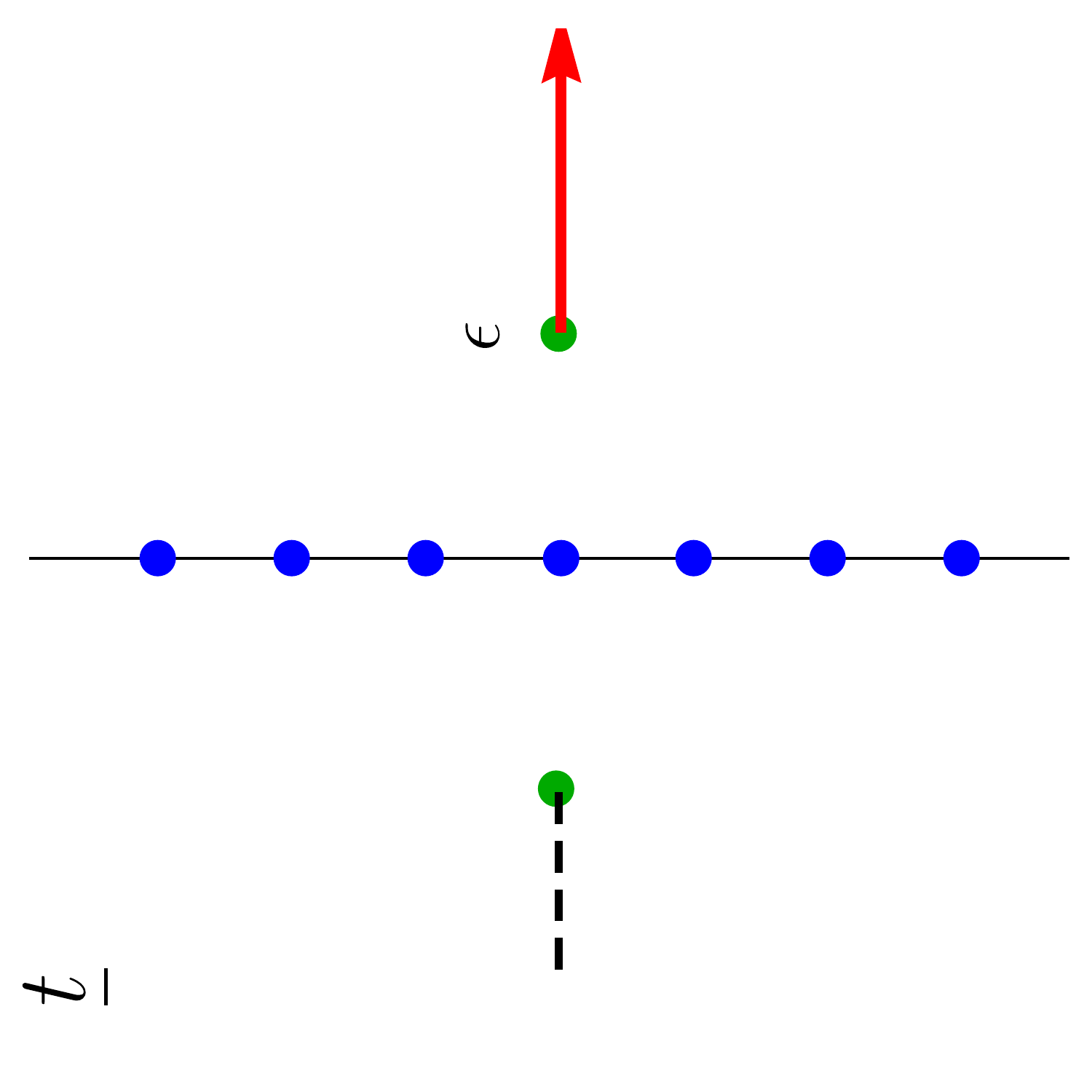}
            \caption[]%
            {{\small \centering  rotated folded contour}}   
        \end{subfigure}
    \caption{We fold the contour $A$ (red) along the branch cut around the branch point $+i \epsilon$ (green dot), and then rotate $u=it$. The blue dots represent the poles of $f(u)$. }\label{fig:hubbard}
\end{figure}

To proceed, we substitute \eqref{eq:NeigvalV}, \eqref{eq:NeigvalS1}, \eqref{eq:NeigvalS2} and use the Hubbard-Stratonovich trick, following the approach in \cite{Anninos:2020hfj,Law:2022zdq}.  For the sum over $L$ in the first line, we can write 
\begin{align}
    \sum_{L\geq0} D_L^3 \, e^{- \left(\lambda^{(V)}_{L,l}+m^2+\frac{1}{\ell_N^2}\right) \tau }=e^{-\frac{\nu_{V,l}^2}{\ell^{2}_N}\tau }\sum_{L=0}^\infty D_{L}^{3}\, e^{-\tau\left(L+\frac12\right)^2/\ell^{2}_N} = e^{-\frac{\nu_{V,l}^2}{\ell^{2}_N} \tau}\int_A du \ \frac{e^{-u^2/4\tau}}{\sqrt{4\pi \tau}} f(u) \; ,
\end{align}
with the integration contour $A = \mathbb{R} + i\delta$, $\delta >0$ (see Fig. \ref{fig:hubbard}). Here we have defined
\begin{equation}
	f(u) \equiv \sum_{L=0}^\infty D_{L}^{3}\, e^{iu\left(L+\frac{1}{2}\right)/\ell_N} = \left(\frac{1+e^{iu/\ell_N}}{1-e^{iu/\ell_N}} \right) \frac{e^{i\frac{u}{2}/\ell_N}}{1-e^{iu/\ell_N}} \; .
\end{equation}
For the sum over $L$ and the last term in the second line of \eqref{eq:narloopstart}, we can write similarly
\begin{align}
    \sum_{L\geq1} D_L^3 \, e^{- \left(\lambda^{(S)}_{L,l}+m^2+\frac{1}{\ell_N^2}\right) \tau }
    =e^{-\frac{\nu_{V,l}^2}{\ell^{2}_N} \tau}\int_A du \ \frac{e^{-u^2/4\tau}}{\sqrt{4\pi \tau}} \left( f(u)- e^{i\frac{u}{2\ell_N}}\right)
\end{align}
and
\begin{align}\label{eq:single}
    -e^{-m^2\tau } = -e^{-\tau\nu_{S,0}^2/\ell^{2}_N } e^{-\tau\left(\frac{1}{2\ell_N}\right)^2}=-e^{-\tau\nu_{S,0}^2/\ell^{2}_N }\int_A du \ \frac{e^{-u^2/4\tau}}{\sqrt{4\pi \tau}}  e^{i\frac{u}{2\ell_N}} \; ,
\end{align}
with the same contour $A$. 
We can then perform the $\tau$-integral in \eqref{eq:narloopstart} (keeping $\Im u = \delta <\epsilon$). Finally, after deforming the contour $A$ as in Fig. \ref{fig:hubbard} and changing variables to $u = it$, we arrive at the regularized formula
\begin{align}
     \log Z_\text{PI} = &\log \widetilde{Z}_{\text{bulk}} - \log Z_{\text{edge}}
\end{align}
where 
\begin{align}\label{eq:narpibulk}
    \log \widetilde{Z}_{\text{bulk}}= \int_\epsilon^\infty& \frac{dt}{2\sqrt{t^2 - \epsilon^2}}\frac{1+q}{1-q}\Bigg[\sum_{l\geq 1} D^d_{l,1}\frac{e^{-\frac{t}{2\ell_N}-i\nu_{V,l}\sqrt{t^2- \epsilon^2}} + e^{-\frac{t}{2\ell_N}+i\nu_{V,l}\sqrt{t^2- \epsilon^2}}}{1-q} \nn\\
    & + \left(\sum_{l\geq 1}+\sum_{l\geq0}\right)D^d_l\frac{e^{-\frac{t}{2\ell_N}-i\nu_{S,l}\sqrt{t^2- \epsilon^2}} + e^{-\frac{t}{2\ell_N}+i\nu_{S,l}\sqrt{t^2- \epsilon^2}}}{1-q}\Bigg]
\end{align}
and
\begin{align}\label{eq:narpiedge}
    \log Z_{\text{edge}}
    =& 2\int_\epsilon^\infty \frac{dt}{2\sqrt{t^2-\epsilon^2}} \sum_{l\geq 0}D^d_l\left(e^{-\frac{t}{2\ell_N}-i\nu_{S,l}\sqrt{t^2- \epsilon^2}} + e^{-\frac{t}{2\ell_N}+i\nu_{S,l}\sqrt{t^2- \epsilon^2}}\right) \; .
\end{align}
Putting $\epsilon=0$, we see that \eqref{eq:narpibulk} recovers \eqref{eq:narrenbulk}, while \eqref{eq:narpiedge} recovers \eqref{eq:narrenedge} plus the term $\log Z^L_{\text{PI}}$. Notice the curious overall factor of 2 in \eqref{eq:narpiedge}, which might be associated with the fact that we have two horizons in the Lorentzian geometry.


\section{Discussion and outlook}\label{sec:discuss}

We have revealed a natural structure for 1-loop Euclidean path integrals of general spinning fields through the relation
\begin{align}\label{coneq:ZPI}
    Z_\text{PI} = \frac{\widetilde{Z}_\text{bulk} }{Z_\text{edge}} \; .
\end{align}
While we have explained in \cite{Law:2022zdq} and generalized in this work the unambiguous canonical meaning of $\widetilde{Z}_\text{bulk}\equiv Z_\text{bulk}/Z^\text{Rin}_\text{bulk}$ as a ratio of thermal canonical partition functions, we have not given the Hilbert space interpretation of $Z_\text{edge}$. 


The presence of the ``edge" contributions is not surprising in view of past studies of entanglement entropy in gauge theories and gravity. In the early work \cite{Kabat:1995eq}, a ``contact term'' was found in the entanglement entropy for Maxwell theory on black holes computed as a conical entropy. A considerable number of works have been devoted to the proper interpretation for such a contact term as ``edge'' degrees of freedom living on the entanglement surface (the bifurcation surface $S^{d-1}$ in the case of black holes). See \cite{Solodukhin:2011gn,Zhitnitsky:2011tr,Donnelly:2011hn,Solodukhin:2012jh,Donnelly:2012st,Casini:2013rba,Donnelly:2014fua,Huang:2014pfa,Donnelly:2015hxa,Casini:2015dsg,Ghosh:2015iwa,Soni:2015yga,Soni:2016ogt,Donnelly:2016auv,Agarwal:2016cir,Blommaert:2018rsf,Blommaert:2018oue,Lin:2018bud,David:2022jfd} for a partial list. While the vast majority of these studies focus on gauge theories and gravity, these edge degrees of freedom are expected to be present for massive spinning fields as well. For one thing, the origin for the contact term in \cite{Kabat:1995eq} is the linear curvature coupling $R$ present in the kinetic term $S\sim\int A \left(-\nabla^2+R \right)A+\dots$, which is also present for massive fields. For another, from the Lorentzian two-sided geometry point of view, the object $Z_\text{bulk}\equiv \Tr \, e^{-\beta_H \hat{H}}$ can be thought of as computing the normalization of the reduced density matrix after tracing out one side. This assumes the global Hilbert space factorizes. For all spinning fields, there are obstructions to this factorization of Hilbert space due to the presence of constraints. In gauge theories and gravity, we have gauge constraints such as the Gauss law constraint $\mathbf{\nabla}\cdot \mathbf{E}=0$ for Maxwell theory; for massive spinning fields, we have for instance the transversality condition $\nabla^\lambda A_\lambda=0$ for a Proca field. The edge degrees of freedom account for the non-factorization of the global Hilbert space due to such constraints. This point has been discussed in \cite{Blommaert:2018rsf} for the Proca field on Rindler space. It is possible that by employing the St\"uckelberg trick, one could understand the massive theory as a gauge theory, so that their edge modes can be understood in the same formalism as in gauge theories and gravity.

In any case, with our purely Euclidean characterization described in Section \ref{sec:dhsspin}, it would be very interesting to connect our work to existing approaches to edge modes and understand their canonical pictures. To that end, we note the crucial role played by the regularity condition imposed on the Euclidean eigenfunctions in the path integral. It is very plausible that the regularity condition is closely related to ``shrinkable boundary condition'' recently discussed by several authors \cite{Donnelly:2018ppr,Jafferis:2019wkd}.

Finally, in the context of holography, our results capture $O(G_N^0)$-effects in the bulk low-energy effective field theory. It would be extremely interesting to investigate the boundary interpretations of the results in \cite{Law:2022zdq} and the current paper. In particular, the boundary counterpart of the edge modes uncovered in this paper might serve as a boundary signature of the bulk black hole horizon. Our results for massive HS on static BTZ could provide a set of concrete data for exploring this direction.

\section*{Acknowledgments}

It is a great pleasure to thank Dionysios Anninos, Frederik Denef, Sean Hartnoll, Daniel Jafferis, and Gabriel Wong for stimulating conversations, and especially Adam Ball and Alejandra Castro for useful discussions and comments on the draft. AL  was supported in part by the Croucher Foundation and the Black Hole Initiative at Harvard University. MG and KP were supported in part by the U.S. Department of Energy grant de-sc0011941.

\appendix

\section{Scalar and vector spherical harmonics on \texorpdfstring{$S^{d-1}$}{}}\label{app:harmonics}


Throughout this paper we use Latin letters such as $i,j,k$ to denote components on $S^{d-1}$. We also use a tilde to denote quantities living intrinsically on $S^{d-1}$; for example, $\tilde \nabla_i$ acts as a covariant derivative with the standard round metric and Levi-Civita connection on $S^{d-1}$.

When $d\geq 4$, we denote the $(d-1)$-dimensional spherical harmonics by $Y_l(\Omega)$, which satisfies
\begin{align}
    -\tilde\nabla^2Y_l=l(l+d-2)Y_l \; , \qquad l\geq 0\; ,
\end{align}
with degeneracy
\begin{align}
    D_l^d=\frac{2l+d-2}{d-2}\binom{l+d-3}{d-3} \; .
\end{align}
Vector spherical harmonics are denoted by $Y_{l,i}(\Omega)$, satisfying
\begin{align}
    -\tilde\nabla^2 Y_{l,i} = \left( l(l+d-2)-1 \right) Y_{l,i} \; , \qquad l\geq 1 \; ,
\end{align}
and the transversality condition
\begin{align}
    \tilde\nabla^i Y_{l,i} = 0 \; ,
\end{align}
with degeneracy 
\begin{align}
    D_{l,1}^d = 
    \frac{l (d+l-2) (d+2 l-2) }{(d-3) (l+1)}\binom{d+l-4}{d-4} \; .
\end{align}

When $d=3$, i.e. on $S^2$ with standard metric $ds^2=d\theta^2 +\sin^2\theta d\varphi^2$, the scalar spherical harmonics are the familiar ones (restoring the magnetic quantum number $m$)
\begin{align}\label{eq:scash2}
    Y_{lm} (\theta,\varphi) \propto e^{im\varphi} P_l^m (\cos\theta) \; , \qquad -l\leq m \leq l\; ,
\end{align}
where $P_l^m(x)$ is the associated Legendre polynomial. The proportionality constant is not important to us. Vector spherical harmonics are related to these by 
\begin{align}\label{appeq:vecsh2}
	Y_{lm, i} (\theta,\varphi)= \frac{1}{\sqrt{l(l+1)}}\epsilon_{ij}\partial^j Y_{lm} (\theta,\varphi) \qquad (l\geq 1)
\end{align}
where $\epsilon_{\theta \phi}= \sin\theta$. Explicitly,
\begin{align}\label{vec SH explicit}
	Y_{lm, \theta} (\theta,\varphi) \propto& \, \sin\theta \, \partial^\varphi Y_{lm} (\theta,\varphi)=\frac{i m}{\sin \theta}Y_{lm} (\theta,\varphi) \nn\\
	Y_{lm, \varphi} (\theta,\varphi)\propto& \, \sin\theta \, \partial^\theta Y_{lm} (\theta,\varphi) \propto \sin^2\theta \, e^{im\varphi}\partial_{x} P_l^m (x)\Big|_{x=\cos\theta} \,.
\end{align}


\section{Scattering in the Rindler-like region}\label{app:rindler}

In this appendix we study massive spinning fields on the Rindler-like wedge:
\begin{align}\label{appeq:refM}
     ds^2 = e^{\frac{4\pi}{\beta}x} \left( -dt^2+dx^2 \right)+r_H^2 d\Omega_{d-1}^2 \; , \qquad -\infty<x<\infty \; .
\end{align}
This is nothing but a product of a 2D Rindler space and a transverse sphere with constant radius $r_H$. Another coordinate system that turns out to be useful is 
\begin{align}\label{appeq:rind new coor}
    y_\pm = \frac{\beta}{2\pi}e^{\mp \frac{2\pi}{\beta} t+\frac{2\pi}{\beta} x} \;,
\end{align}
in terms of which the metric becomes 
\begin{align}\label{appeq:newrindmetric}
    ds^2 &= dy_+ dy_-  + r_H^2 d\Omega_{d-1}^2  \; .
\end{align}

\subsection{Massive scalar}

This case has been studied in the Appendix A of \cite{Law:2022zdq}. The Klein-Gordon equation $(-\nabla^2+m^2)\, \phi=0$ is equivalent to
\begin{align}\label{appeq:rindeqs0}
	\left[ -\partial_t^2 +\partial_x^2 +e^{\frac{4\pi}{\beta}x}\left( \frac{\tilde\nabla^2}{r_H^2}-m^2\right)  \right] \phi =0 \;.
\end{align}
Solving with the ansatz
\begin{align}
    \phi(t,x,\Omega)=e^{-i \omega t} \psi(x)Y_l(\Omega)\;,
\end{align}
the normalizable solution is
\begin{align}\label{appeq:rins0sol}
    \psi^\text{Scalar}_{\omega l} (x') = K_{\frac{i \beta  \omega }{2 \pi }}\left(2 e^{\frac{2 \pi  x'}{\beta }}\right) \; ,
\end{align}
where
\begin{align}
    x' = x + \frac{\beta}{2\pi} \log \frac{\beta M_{l}}{4\pi} \; ,\qquad M_{l} \equiv \sqrt{\frac{l(l+d-2)}{r_H^2}+m^2} \; .
\end{align}
Near the horizon \eqref{appeq:rins0sol} takes the asymptotic form
\begin{align}\label{appeq:s0nearhor}
   \psi^\text{Scalar}_{\omega l} (x'\to -\infty)  \propto \Gamma \left(\frac{i
   \beta  \omega }{2 \pi }\right) e^{-i  \omega x'}+ \Gamma \left(-\frac{i \beta  \omega }{2 \pi }\right)
   e^{i  \omega x'}\; .
\end{align}
The ratio between the coefficients of the outgoing and incoming waves
\begin{align}\label{appeq:rinseedS}
    \mathcal{S}^{\text{Rin}}(\beta,\omega) = \frac{\Gamma
   \left(\frac{i \beta  \omega }{2 \pi }\right)}{\Gamma
   \left(-\frac{i \beta  \omega }{2 \pi }\right)}
\end{align}
is a pure phase, or rank-1 unitary S-matrix.

\subsection{Massive vector in any \texorpdfstring{$d\geq 3$}{}}

In this section we consider a vector field of mass $m^2$ on the wedge \eqref{appeq:refM}. In the coordinates \eqref{appeq:newrindmetric}, all Christoffel symbols except $\Gamma^i_{jk}$ are trivial, and one immediately concludes that the Proca equation of motion $\nabla^\mu F_{\mu \nu} = m^2 A_\nu$ is equivalent to
\begin{align}\label{appeq:rindeq}
	\left[ -\partial_t^2 +\partial_x^2 +e^{\frac{4\pi}{\beta}x}\left( \frac{\tilde\nabla^2}{r_H^2}-m^2\right)  \right] A_\pm =0 =
	\left[ -\partial_t^2 +\partial_x^2 +e^{\frac{4\pi}{\beta}x}\left( \frac{\tilde\nabla^2-(d-2)}{r_H^2}-m^2\right)  \right] A_i  \;,
\end{align}
together with the transversality condition
\begin{align}\label{appeq:transrinpm}
    2\left(\partial_- A_+ + \partial_+ A_- \right)+  \frac{1}{r_H^2} \tilde\nabla^i A_i = 0 \; ,
\end{align}
Here we have abbreviated $\partial_\pm \equiv \partial_{y_\pm}$. Note $A_\pm$ are related to $A_t,A_x$ through
\begin{align}\label{appeq:rindpm}
    A_\pm= e^{\pm\frac{2\pi}{\beta}t -\frac{2\pi}{\beta} x}\left(\mp A_t + A_x \right) \; .
\end{align}

Since all equations in \eqref{appeq:rindeq} take the same form as the scalar case \eqref{appeq:rindeqs0}, we can immediately employ the results from the last section, except that there can be different types of solutions according to $SO(d)$ irreducible representations. 

\paragraph{Vector type}

First we have the vector-type solutions of the form
\begin{align}\label{appeq:rinvecan}
    A_\pm = 0 \; , \qquad 
    A_i = e^{-i \omega t}\psi^{(V)}_l(x) \,Y_{l,i}(\Omega)  \; ,
\end{align}
where the angular dependence of $A_i$ are taken to be vector spherical harmonics $Y_{l,i}(\Omega)$. For this type of solutions, the transversality condition \eqref{appeq:transrinpm} is trivially satisfied. The only non-trivial equation in \eqref{appeq:rindeq} with ansatz \eqref{appeq:rinvecan} is 
\begin{align}\label{app:rindvecscatt}
	\left[-\partial_{x'}^2 + \left(\frac{4\pi}{\beta}\right)^2 e^{\frac{4\pi}{\beta}x'} \right] \psi^{(V)}_l(x')  = \omega^2 \psi^{(V)}_l(x')\;, \qquad 
\end{align}
Here we have defined 
\begin{align}
    x' = x + \frac{\beta}{2\pi} \log \frac{\beta M_{1,l}}{4\pi} \; ,\qquad M_{1,l} \equiv \sqrt{\frac{(l+1)(l+d-3)}{r_H^2}+m^2} \; .
\end{align}
The normalizable solution to \eqref{app:rindvecscatt} is
\begin{align}
    \psi^{(V)}_{\omega l}(x')  =\psi^\text{Scalar}_{\omega l} (x') = K_{\frac{i \beta  \omega }{2 \pi }}\left(2 \, e^{\frac{2 \pi  x'}{\beta }}\right) \;,
\end{align}
and the S-matrix is simply
\begin{align}
    \mathcal{S}^{(\text{Rin},V)}(\beta,\omega) =\mathcal{S}^{\text{Rin}}(\beta,\omega)= \frac{\Gamma
   \left(\frac{i \beta  \omega }{2 \pi }\right)}{\Gamma
   \left(-\frac{i \beta  \omega }{2 \pi }\right)} \; .
\end{align}

\paragraph{Scalar type}

Next we have the scalar-type solutions, with ansatz
\begin{align}\label{appeq:rindscaan}
     A_\pm = \, e^{\pm \frac{2\pi}{\beta} t-i\omega t} \psi_{\pm,l}^{S} (x)\, Y_{l}(\Omega) \; , \qquad A_i = \, e^{-i\omega t} \psi_{i,l}^{S} (x)\, \frac{\tilde\nabla_i Y_{l}(\Omega)}{l(l+d-2)} \; .
\end{align}
Here we have inserted a factor $e^{\pm \frac{2\pi}{\beta} t}$ to compensate for the corresponding factors in \eqref{appeq:rindpm} to get a normal mode with time dependence $e^{-i\omega t}$. We will focus on solving for $\psi_{\pm,l}^{S}$, with which the angular solution $\psi_{i,l}^{S}$ is completely determined through \eqref{appeq:transrinpm}. For each $l\geq 1$, there are two linearly independent solutions, which can be taken to be
\begin{align}\label{appeq:pmtype}
    \text{Scalar $\pm$-type: }& \qquad A_\pm \neq 0 \; , \qquad A_\mp =0 \; , \qquad A_i \neq 0  \; .
\end{align}
For the $\pm$-solution with $A_\mp = 0$, plugging the ansatz \eqref{appeq:rindscaan} into \eqref{appeq:rindeq} leads to 
\begin{align}\label{appeq:rindeqsca1}
	\left[-\partial_{x'}^2 + \left(\frac{4\pi}{\beta}\right)^2 e^{\frac{4\pi}{\beta}x'} \right] \psi_{\pm,l}^{S,\pm} (x')& = \left(\omega\pm i\frac{2\pi}{\beta}\right)^2 \psi_{\pm,l}^{S,\pm}(x') \nn\\
	\left[-\partial_{x'}^2 + \left(\frac{4\pi}{\beta}\right)^2 e^{\frac{4\pi}{\beta}x'} \right] \psi_{i,l}^{S,\pm}(x') &=\omega^2 \psi_{i,l}^{S,\pm}(x')\;.
\end{align}
Here we have defined 
\begin{align}
    x' = x + \frac{\beta}{2\pi} \log \frac{\beta M_{0,l}}{4\pi} \; ,\qquad M_{0,l} \equiv \sqrt{\frac{l(l+d-2)}{r_H^2}+m^2} \; .
\end{align}
The normalizable solutions are
\begin{align}\label{appeq:tower1sol}
    \psi_{\pm,l}^{S,\pm}(x') = \psi^\text{Scalar}_{\omega\pm i \frac{2\pi}{\beta}, l} (x')= K_{\frac{i \beta  \omega }{2 \pi }\mp 1}\left(2\, e^{\frac{2 \pi  x'}{\beta }}\right) \; , 
    \qquad (l\geq 1) \; .
\end{align}
Using 
\begin{align}\label{appeq:hankelder}
     \frac{d}{dz} K_\alpha(z) =\pm\frac{\alpha}{z} K_\alpha(z) - K_{\alpha\pm 1}(z) \;, 
\end{align}
one can check that the angular solutions $\psi_{i,l}^{S}$ obtained through \eqref{appeq:transrinpm} automatically solve the angular equation \eqref{appeq:rindeqsca1}. Since the $\pm$-modes \eqref{appeq:tower1sol} are same as \eqref{appeq:rins0sol} but with $\omega \to \omega \pm i\frac{2\pi}{\beta}$, we can immediately write down the S-matrices
\begin{align}\label{appeq:pmSmat}
    \mathcal{S}_{l\geq 1}^{\text{Rin},\pm}(\beta,\omega) = \mathcal{S}^{\text{Rin}}\left(\beta,\omega\pm i\frac{2\pi}{\beta} \right) \; , 
\end{align}
with $\mathcal{S}^{\text{Rin}}(\beta,\omega)$ defined in \eqref{appeq:rinseedS}. Since the equations \eqref{appeq:rindeqsca1} for $\psi_{\pm,l}^{S,\pm}$ are related through $\omega \to -\omega$, they can be thought of as a time-reversal dual pair of scattering problems, and their S-matrices \eqref{appeq:pmSmat} satisfy the unitary condition
\begin{align}
    \mathcal{S}^{\text{Rin},+}(\beta,-\omega)\, \mathcal{S}^{\text{Rin},-}(\beta,\omega) =1 \; .
\end{align}

When $l=0$, the ansatz for $A_i$ in \eqref{appeq:rindscaan} breaks down, and we have instead 
\begin{align}\label{appeq:rindscaan0}
     A_\pm =  e^{\pm \frac{2\pi}{\beta} t-i\omega t} \psi_{\pm,l=0}^{S} \,  \; , \qquad A_i = 0 \; ,
\end{align}
where both $A_\pm$ must be non-zero. We still have the first line of \eqref{appeq:rindeqsca1}, with normalizable solutions
\begin{align}
    \psi_{\pm,l=0}^{S}(x') = \psi^\text{Scalar}_{\omega\pm i \frac{2\pi}{\beta}, l=0} (x')= K_{\frac{i \beta  \omega }{2 \pi }\mp 1}\left(2\, e^{\frac{2 \pi  x'}{\beta }}\right) \;  ,\qquad (l=0) \; .
\end{align}
The transverality condition \eqref{appeq:rindpm} fixes the relative coefficients of these two solutions, so that the full solution is
\begin{align}\label{appeq:tower1sol0}
    \left(A_+,A_-\right)_{l=0} = e^{-i\omega t} \left( e^{\frac{2\pi}{\beta} t} \psi_{+,l=0}^{S}\,,\,-e^{- \frac{2\pi}{\beta} t} \psi_{-,l=0}^{S}\right) \;.
\end{align}
Near horizon, this behaves as
\begin{equation}\begin{split}
    \left(A_+, A_- \right)_{ l=0} (x'\to -\infty) \;
    \propto  &\;\; \Gamma \left( \frac{i \beta \omega}{2\pi}+1\right)  (0,1)\,e^{\left(-i\omega-\frac{2\pi}{\beta} \right) (t+x')}\\ &+\Gamma \left( -\frac{i \beta \omega}{2\pi}+1\right)  (-1,0)\, e^{\left(-i\omega+\frac{2\pi}{\beta} \right)(t-x')} \;.
    \end{split}
\end{equation}
Here the second and first terms correspond to waves incoming from and outgoing to the horizon respectively, the ratio of their coefficients again defines a unitary S-matrix:
\begin{align}\label{appeq:rindl0S}
    \mathcal{S}_{l=0}^{\text{Rin}}(\beta,\omega) = \frac{\Gamma \left( \frac{i \beta \omega}{2\pi}+1\right) }{\Gamma \left( -\frac{i \beta \omega}{2\pi}+1\right) } \;.
\end{align}

\subsection{Massive higher spin in \texorpdfstring{$d=2$}{}}\label{app:rindler3D}

When $d=2$, the transverse sphere becomes a circle $S^1$ with radius $r_H$, and the metric is simply 
\begin{align}\label{appeq:rin3D}
     ds^2 = e^{\frac{4\pi}{\beta}x} \left( -dt^2+dx^2 \right)+r_H^2 d\vartheta^2 \; , \qquad -\infty<x<\infty \; ,\qquad \vartheta \simeq\vartheta+2\pi \; .
\end{align}

\subsubsection{Massive vector}

Now let us study a massive vector on \eqref{appeq:rin3D}. A special feature for the case of $d=2$ is that \eqref{appeq:rindeq} and \eqref{appeq:transrinpm} can be equivalently described by either $(\mp)$ set of first-order equations
\begin{align}\label{appeq:rin1sts1}
    \epsilon_{\alpha\mu\nu}\nabla^\mu A^\nu=\mp m A_\alpha \;. 
\end{align}
Here we take the following convention for the Levi-Civita symbol
\begin{align}
	\epsilon^{\mu\nu\lambda}\equiv \frac{\tilde{\epsilon}^{\mu\nu\lambda}}{\sqrt{-g}} \; , \qquad \tilde{\epsilon}^{-+\vartheta} = -\tilde{\epsilon}_{-+\vartheta} =1 \; .
\end{align} 
It is straightforward to check that solutions to \eqref{appeq:rin1sts1} satisfy \eqref{appeq:rindeq} and \eqref{appeq:transrinpm} at $d=2$. In a parity-invariant theory, both sets of solutions should be included. To obtain the explicit solutions, note that equations \eqref{appeq:rindeq} and \eqref{appeq:transrinpm} remain valid. The only difference is that the isometry group for $S^1$ is $U(1)$, and we do not have to separate the solutions into vector or scalar type. The full set of normal mode solutions is obtained using the ansatz
\begin{align}\label{appeq:rindscaand2}
     A_\pm = C_\pm \, e^{\pm \frac{2\pi}{\beta} t-i\omega t} \psi_{\pm,l}(x)\, e^{il\vartheta} \; , \qquad A_\vartheta = C_\vartheta \, e^{-i\omega t} \psi_{\vartheta,l} (x)\, e^{il\vartheta}\; , \qquad l=0,\pm 1,\pm 2,\dots  \; .
\end{align}
Here we have inserted overall constants $C_\pm$ and $C_\vartheta$ to be determined below. Notice that the $U(1)$ angular momentum $l$ takes values over all integers. For each $l\in \mathbb{Z}$, we obtain
\begin{align}\label{appeq:tower1sold2}
    \psi_{\pm,l}(x') = \psi^\text{Scalar}_{\omega\pm i \frac{2\pi}{\beta}, l} (x') \; , 
    \qquad
    \psi_{\vartheta,l} (x') = \psi^\text{Scalar}_{\omega, l} (x')
    \qquad l\in \mathbb{Z} \; ,
\end{align}
where 
\begin{align}\label{appeq:d2x}
    x' = x + \frac{\beta}{2\pi} \log \frac{\beta M_{0,l}}{4\pi} \; ,\qquad M_{0,l} \equiv \sqrt{\frac{l^2}{r_H^2}+m^2} \; .
\end{align}

\paragraph{Polarization vectors}

In the first-order formulation \eqref{appeq:rin1sts1}, all components \eqref{appeq:tower1sold2} are coupled. To find their relative coefficients, we note that the $\vartheta$-component of \eqref{appeq:rin1sts1} reads explicitly
\begin{align}\label{appeq:rin1stex}
    2r_H \left( \partial_+ A_- -\partial_- A_+ \right)=\pm m A_\vartheta \; .
\end{align}
On the other hand, the transversality condition \eqref{appeq:transrinpm} at $d=2$ implies for the ansatz \eqref{appeq:rindscaand2}
\begin{align}\label{appeq:rintransex}
    i l A_\vartheta = -2 r_H^2 \left( \partial_+ A_- +\partial_- A_+ \right) \;. 
\end{align}
These then lead to the relation
\begin{align}
    \left(i l \pm m r_H \right) \partial_+ A_- = \left(i l \mp m r_H \right) \partial_- A_+ \;,
\end{align}
where the upper (lower) signs correspond to the $-$ ($+$)-branch \eqref{appeq:rin1sts1}. Plugging in \eqref{appeq:rindscaand2} and \eqref{appeq:tower1sold2}, one finds
\begin{align}
    C^{(\mp)}_{l,-} \left(i l \pm m r_H \right)  = C^{(\mp)}_{l,+} \left(i l \mp m r_H \right)  \;. 
\end{align}
To summarize, we have determined the two sets of normal modes
\begin{align}\label{appeq:rinvec1stsol}
    \left(A_+, A_- \right)^{(\mp)}_{\omega l} = e^{ - i \omega t+i l\vartheta} \left(C^{(\mp)}_{l,+} \, e^{ \frac{2\pi}{\beta} t} \psi_{\pm,l}(x') , C^{(\mp)}_{l,-} \, e^{- \frac{2\pi}{\beta} t} \psi_{-,l}(x') \right)
\end{align}
satisfying the $\mp$ equations \eqref{appeq:rin1sts1} respectively, with $A_\vartheta$ uniquely determined by \eqref{appeq:rin1stex} or \eqref{appeq:rintransex}.

\paragraph{S-matrices}

Comparing \eqref{appeq:tower1sold2} and \eqref{appeq:s0nearhor}, we see that near horizon the incoming and outgoing waves are dominated by the components $A_+$ and $A_-$ respectively. Explicitly,
\begin{align}\label{appeq:rinvec1stsolnh}
    \left(A_+, A_- \right)^{(\mp)}_{\omega l} (x'\to -\infty) 
    \propto  \, B^{(\mp),\text{out}}_{\omega l}\left(0,1 \right)e^{ - i \left(\omega-i\frac{2\pi}{\beta}\right) (t+x')}+B^{(\mp),\text{in}}_{\omega l}\left(1,0 \right)e^{ - i \left(\omega+i\frac{2\pi}{\beta}\right) (t-x')} \;.
\end{align}
We have suppressed the $e^{ i l\vartheta}$ dependence. The outgoing and incoming coefficients are
\begin{align}
    B^{(\mp),\text{out}}_{\omega l}=\left(i l \mp m r_H \right) \Gamma\left(\frac{i\beta \omega}{2\pi} +1\right) \; , \qquad B^{(\mp),\text{in}}_{\omega l} = \left(i l \pm m r_H \right) \Gamma\left(-\frac{i\beta \omega}{2\pi} +1\right)\;,
\end{align}
whose ratio 
\begin{align}\label{appeq:rins1Smat}
    \mathcal{S}_{l}^{\text{Rin}, (\pm, s=1)} (\beta,\omega)= \frac{B^{(\mp),\text{out}}_{\omega l}}{B^{(\mp),\text{in}}_{\omega l}}=\frac{i l \mp m r_H}{i l \pm m r_H}\frac{\Gamma\left(\frac{i\beta \omega}{2\pi} +1\right)}{\Gamma\left(-\frac{i\beta \omega}{2\pi} +1\right)} \equiv \frac{i l \mp m r_H}{i l \pm m r_H}\mathcal{S}^{\text{Rin}, ( s=1)}(\omega)
\end{align}
defines a unitary S-matrix.



\subsubsection{Massive higher spin}

We now study a general spin-$s$ symmetric tensor field $\phi_{\mu_1 \mu_2 \cdots \mu_s}$ with mass $m^2$, described by either $(\mp)$ set of first-order equations
\begin{align}\label{appeq:rinhs1st}
	\epsilon\indices{_{\mu_1}^\alpha^\beta}\nabla_\alpha \phi_{\beta \mu_2 \cdots \mu_s} = \mp m\, \phi_{\mu_1 \mu_2 \cdots \mu_s} \; .
\end{align}
One can show that the solutions to these equations solve the Fierz-Pauli system
\begin{align}\label{appeq:rinFP}
    \left(-\nabla^2+m^2 \right)\phi_{\mu_1 \mu_2 \cdots \mu_s} =0\;, \qquad \nabla^\lambda \phi_{\lambda \mu_1 \mu_2 \cdots \mu_{s-1}} =0 \; , \qquad \phi\indices{^\lambda_{\lambda \mu_1 \mu_2 \cdots \mu_{s-2}}}=0\; .
\end{align}
In the coordinates \eqref{appeq:rin3D}, all Christoffel symbols are trivial, and one immediately concludes that the components with respect to \eqref{appeq:newrindmetric} all satisfy the scalar equation
\begin{align}
    \left[ -\partial_t^2 +\partial_x^2 +e^{\frac{4\pi}{\beta}x}\left( \frac{\tilde\nabla^2}{r_H^2}-m^2\right)  \right] \phi_{A_1 A_2 \cdots A_s} =0 \; ,
\end{align}
where $A_I \in (\pm,\vartheta)$. Because of this, the normal mode functions are all given by the scalar mode function \eqref{appeq:rins0sol} (with $x'$ defined in \eqref{appeq:d2x}), with appropriate (complex) shifts in $\omega$ dictated by the relation between the $(+,-,\vartheta)$-components and the original $(t,x,\vartheta)$-components. For example, for a spin-2 field $h_{\mu\nu}$, 
\begin{align}
    h_{tt}=& \, e^{-\frac{4\pi}{\beta}(t-x)}h_{++}-2\, e^{\frac{4\pi}{\beta}x}h_{+-}+\, e^{\frac{4\pi}{\beta}(t+x)}h_{--}\nn\\
    h_{tx}=& -2\, e^{-\frac{4\pi}{\beta}(t-x)}h_{++}+2\, e^{\frac{4\pi}{\beta}(t+x)}h_{--}\nn\\
    h_{xx}=& \, e^{-\frac{4\pi}{\beta}(t-x)}h_{++}+2\, e^{\frac{4\pi}{\beta}x}h_{+-}+\, e^{\frac{4\pi}{\beta}(t+x)}h_{--}\nn\\
    h_{t\vartheta}=&- \, e^{-\frac{2\pi}{\beta}(t-x)}h_{+\vartheta}+\, e^{\frac{2\pi}{\beta}(t+x)}h_{-\vartheta}\nn\\
    h_{x\vartheta}=& \, e^{-\frac{2\pi}{\beta}(t-x)}h_{+\vartheta}+\, e^{\frac{2\pi}{\beta}(t+x)}h_{-\vartheta}\;,
\end{align}
which imply that the normal modes are solved with the ansatz
\begin{gather}
    h_{\pm \pm} = \, e^{\pm \frac{4\pi}{\beta} t-i\omega t+il\vartheta} \psi_{\pm\pm,l} (x) \; , \qquad h_{+-} = \, e^{-i\omega t+il\vartheta} \psi_{+-,l} (x) \; , \nn\\
      \qquad 
    h_{\pm \vartheta} = \, e^{\pm \frac{2\pi}{\beta} t-i\omega t+il\vartheta} \psi_{\pm\vartheta,l} (x) \; ,\qquad h_{\vartheta\vartheta} = \, e^{-i\omega t+il\vartheta} \psi_{\vartheta\vartheta,l} (x)  \;,
\end{gather}
where $l\in \mathbb{Z}$. We can immediately write down the explicit solutions:
\begin{gather}
    \psi_{\pm\pm,\omega l} (x)  = \psi^\text{Scalar}_{\omega\pm 2 i \frac{2\pi}{\beta}, l} (x') =K_{\frac{i \beta  \omega }{2 \pi }\mp 2}\left(2\, e^{\frac{2 \pi  x'}{\beta }}\right) \;,\nn\\
    \psi_{\pm\vartheta,\omega l} (x) = \psi^\text{Scalar}_{\omega\pm i \frac{2\pi}{\beta}, l} (x')= K_{\frac{i \beta  \omega }{2 \pi }\mp 1}\left(2\, e^{\frac{2 \pi  x'}{\beta }}\right) \;, \nn\\ \psi_{+-,\omega l} (x) =\psi_{\vartheta\vartheta,l} (x)= \psi^\text{Scalar}_{\omega l} (x')=K_{\frac{i \beta  \omega }{2 \pi }}\left(2\, e^{\frac{2 \pi  x'}{\beta }}\right) \;.
\end{gather}
It is straightforward to generalize to arbitrary spin $s\geq 2$. We use the notation $\phi_{(a)(b)(c)}$ to denote the component of a spin-$s$ field with $a$ $+$-, $b$ $-$-, and $c=s-a-b$ $\vartheta$-indices, i.e.
\begin{align}\label{appeq:hsnotation}
    \phi_{(a)(b)(c)} \equiv \phi_{\underbrace{+\cdots +}_{a}\underbrace{-\cdots -}_{b}\underbrace{\vartheta\cdots \vartheta}_{c}} \; .
\end{align}
Normal modes are solved with the ansatz
\begin{align}\label{appeq:rinhsan}
    \phi_{(a)(b)(c)} =C_{(a)(b)(c)} e^{(a-b)\frac{2\pi}{\beta}t-i\omega t}\psi_{(a)(b)(c),\omega l}(x)\, e^{il\vartheta}\;, \qquad l \in\mathbb{Z}\; ,
\end{align}
where we have inserted the relative coefficients $C_{(a)(b)(c)}$ to be determined. The solutions are then
\begin{align}\label{appeq:rinhsd2sol}
    \psi_{(a)(b)(c),\omega l}(x)= \psi^\text{Scalar}_{\omega+ (a-b)i\frac{2\pi}{\beta}, l} (x') = K_{\frac{i \beta  \omega }{2 \pi }+ (b-a)}\left(2\, e^{\frac{2 \pi  x'}{\beta }}\right) \;.
\end{align}

\paragraph{Polarization tensors}

In the first-order formulation \eqref{appeq:rinhs1st}, all components \eqref{appeq:rinhsan} are coupled. To find their relative coefficients, we note that the $\vartheta$-component of \eqref{appeq:rinhs1st} reads explicitly
\begin{align}\label{appeq:rin1stexhs}
    2r_H \left( \partial_+ \phi_{(a-1)(b+1)(c)} -\partial_- \phi_{(a)(b)(c)} \right)=\pm m\, \phi_{(a-1)(b)(c+1)} \; .
\end{align}
On the other hand, the transversality condition \eqref{appeq:transrinpm} at $d=2$ implies for the ansatz \eqref{appeq:rindscaand2}
\begin{align}\label{appeq:rintransexhs}
    il \phi_{(a-1)(b)(c+1)} = -2 r_H^2 \left( \partial_+ \phi_{(a-1)(b+1)(c)} +\partial_- \phi_{(a)(b)(c)} \right) \;. 
\end{align}
These then lead to the relation
\begin{align}
    \left(i l \pm m r_H \right) \partial_+ \phi_{(a-1)(b+1)(c)} = \left(i l \mp m r_H \right) \partial_- \phi_{(a)(b)(c)} \;,
\end{align}
where the upper (lower) signs correspond to the $-$ ($+$)-branch \eqref{appeq:rinhs1st}. Plugging in \eqref{appeq:rinhsan} and \eqref{appeq:rinhsd2sol}, one then finds
\begin{align}
    C^{(\mp)}_{l,(a-1)(b+1)(c)} \left(i l \pm m r_H \right)  = C^{(\mp)}_{l,(a)(b)(c)} \left(i l \mp m r_H \right)  \;. 
\end{align}

\paragraph{S-matrices}

Analogous to the vector case, near horizon the incoming and outgoing waves are dominated by the components $\phi_{(s)(0)(0)}$ and $\phi_{(0)(s)(0)}$ respectively, and we have
\begin{align}
    &\left(\phi_{(s)(0)(0)}, \phi_{(0)(s)(0)} \right)^{(\mp)}_{\omega l}  (x\to -\infty) \nn\\
    \propto & \, B^{(\mp),\text{out}}_{\omega l}\left(0,1 \right)e^{ - i \left(\omega-is\frac{2\pi}{\beta}\right) (t+x')}+B^{(\mp),\text{in}}_{\omega l}\left(1,0 \right)e^{ - i \left(\omega+is\frac{2\pi}{\beta}\right) (t-x')} \;,
\end{align}
where we have suppressed the $e^{ i l\vartheta}$ dependence, and the outgoing and incoming coefficients are
\begin{align}
    B^{(\mp),\text{out}}_{\omega l}=\left(i l \mp m r_H \right)^s \Gamma\left(\frac{i\beta \omega}{2\pi} +s\right) \; , \qquad B^{(\mp),\text{in}}_{\omega l} = \left(i l \pm m r_H \right)^s \Gamma\left(-\frac{i\beta \omega}{2\pi} +s\right)\;,
\end{align}
whose ratio 
\begin{align}\label{appeq:rinhsSmat}
    \mathcal{S}_{ l}^{\text{Rin}, (\pm, s)} (\beta,\omega)= \frac{B^{(\mp),\text{out}}_{\omega l}}{B^{(\mp),\text{in}}_{\omega l}}= \left(\frac{i l \mp m r_H}{i l \pm m r_H} \right)^s\frac{\Gamma\left(\frac{i\beta \omega}{2\pi} +s\right)}{\Gamma\left(-\frac{i\beta \omega}{2\pi} +s\right)} =\left(\frac{i l \mp m r_H}{i l \pm m r_H} \right)^s\mathcal{S}^{\text{Rin}, ( s)} (\beta,\omega)
\end{align}
defines a unitary S-matrix. Notice that the overall factor is independent of $\omega$ and drops out in relation \eqref{eq:rhochange}.

\section{Massive higher spin on global \texorpdfstring{$AdS_3$}{}}\label{app:tads}

Even though global $AdS_3$ (setting $\ell_\text{AdS}=1$)
\begin{align}\label{appeq:ads3met}
    ds^2 = -\left(1+r^2 \right) dt^2 + \frac{dr^2}{1+r^2 } +r^2 d\vartheta^2 
\end{align}
does not have a horizon and the considerations in the main text do not apply, we include this example due to its relation with the BTZ case. Also, it is instructive to highlight the difference between the two computations.

\paragraph{Thermal canonical partition function}

The {\it normal} mode spectrum for a field with spin $s\geq 1$ and generic mass $m^2= (\Delta -s)(\Delta+s-2)$ on global $AdS_{3}$ is well-known:
\begin{align}\label{eq:adsnormal}
	\omega_{nl}= 2n+|l|+\Delta
\end{align}
where $n=0,1,2,\dots$ and $l=0,\pm 1,\pm 2, \dots$ labels the $U(1)$ angular momentum quantum number. In this case the density of state is simply a sum of delta functions over the discrete spectrum \eqref{eq:adsnormal}. The thermal canonical partition function is
\begin{align}
\log Z_\text{bulk}^{AdS_3}\equiv \log \Tr \, e^{-\beta \hat{H}} =-2\sum_{n,l} \log (1-e^{-\omega_{nl}\beta}) \;.
\end{align}
Here we have dropped an infinite contribution from zero point energies that renormalizes the cosmological constant. Expanding the logarithm and performing the sums over $n,l$, we have
\begin{align}\label{appeq:TAdS}
	\log Z_\text{bulk}^{AdS_3}=\sum_{k=1}^\infty \frac{2}{k}\frac{e^{-\Delta k \beta}}{(1-e^{-k \beta})^2}=\sum_{k=1}^\infty \frac{\chi_{[\Delta,s]}^{AdS_{3}}(k \beta)}{k}\; , \qquad \chi_{[\Delta,s]}^{AdS_{3}}(t)=2\frac{e^{-\Delta t}}{(1-e^{-t})^2}\;.
\end{align}
In the last equality we have expressed the result in terms of the $SO(2,2)$ group character $\chi_{[\Delta,s]}^{AdS_{3}}(t)$.

\paragraph{Path integral on thermal $AdS$ ($TAdS$)}

The same result can be obtained by computing the Euclidean path integral on $TAdS_3$: 
\begin{align}\label{appeq:scalar tads pi}
	\log Z_\text{PI}^{TAdS_3} =\int_0^\infty \frac{d\tau}{2\tau} e^{-\epsilon^2/4\tau} \Tr \, e^{-\left(-\nabla_s^2+M_s^2\right) \tau}
\end{align}
where $e^{-\epsilon^2/4\tau}$ is a UV regulator. Here the trace Tr is over the spectrum of the Laplace operator $-\nabla_s^2+M_s^2$ on $TAdS_3$. This has been computed by the image method in \cite{David:2009xg}:\footnote{The heat kernel for $s\leq 2$ on $TAdS_3$ was first computed in \cite{Giombi:2008vd}.}
\begin{align}
	\Tr \, e^{-\left(-\nabla_s^2+M_s^2\right) \tau} =  \sum_{k=1}^\infty \frac{\beta}{\sqrt{4\pi \tau} \sinh^2\frac{k\beta}{2}} e^{-\frac{k\beta^2}{4\tau}}e^{-(\Delta-1)^2 \tau} \; .
\end{align}
Performing the $\tau$-integral in \eqref{appeq:scalar tads pi} and putting $\epsilon=0$, we recover the canonical result \eqref{appeq:TAdS}.

\section{Massive higher spin on BTZ: normal modes}\label{app:btz}

In this appendix we find explicitly the normal mode solutions for massive higher spin (HS) fields on a static BTZ black hole (setting $\ell_\text{AdS}=1$):
\begin{equation}\label{appeq:btzmetric}
	ds^2 = \frac{r_H^2}{\sinh^2 (r_H x)} \left( -dt^2 + dx^2 + \cosh^2 (r_H x) d\vartheta^2 \right) \;, \qquad -\infty < x<0\; .
\end{equation}
Another coordinate system that turns out to be useful is given by
\begin{align}\label{appeq:new coor}
    y_\pm = e^{\mp r_H t} \sech (r_H x) \;,
\end{align}
in terms of which the metric becomes 
\begin{align}\label{appeq:newbtzmetric}
    ds^2 &= \frac{1}{4(1- y_+ y_-)^2} \Big( y_-^2 dy_+^2 + 2(2-y_+y_-) dy_+dy_- + y_+^2 dy_-^2 \Big) + \frac{r_H^2}{1-y_+y_-} d\vartheta^2 \,.
\end{align}
When expressing quantities in these coordinates, we use the shorthand notations $\pm$ to denote the $y_\pm$-components. The non-zero Christoffel symbols associated with the metric \eqref{appeq:newbtzmetric} are
\begin{align}
    \Gamma^\pm_{\pm\pm}=\frac{y_\mp}{1-y_+y_-} \; , \qquad  \Gamma^\pm_{\pm\mp}=\frac{1}{2}\frac{y_\pm}{1-y_+y_-}\; , \qquad \Gamma^\pm_{\vartheta\vartheta} = -r_H^2 y_\pm  \; , \qquad \Gamma^\vartheta_{\pm \vartheta}=\frac{1}{2}\frac{y_\mp}{1-y_+y_-} \;, 
\end{align}
which satisfy
\begin{align}
    \Gamma^\pm_{\pm\pm}= 2 \Gamma^\vartheta_{\pm \vartheta}=2\Gamma^\mp_{\pm\mp} \;.
\end{align}



\subsection{Massive scalars}

For a scalar with mass $m^2 = \Delta(\Delta-2)$, rescaling 
\begin{align}\label{appeq:btzscresc}
    \phi(t,x,\vartheta) =\sqrt{-\tanh (r_H x)}\, \psi(t,x,\vartheta)\;,
\end{align}
the Klein-Gordon equation $\left( -\nabla^2+m^2\right)\phi=0$ becomes
\begin{align}\label{appeq:btzscalar}
    \left[ - \partial_t^2 + \partial_x^2 + \frac{r_H^2}{\sinh^2(2 r_H x)} -\frac{(\Delta -1)^2 r_H^2}{\sinh^2 (r_H x)}  + \frac{\partial_\vartheta^2}{\cosh^2(r_H x)}\right] \psi(x) = 0 \;.
\end{align}
Solving with the ansatz
\begin{align}\label{appeq:btzscans}
    \psi(t,x,\vartheta)= e^{-i\omega t+i l \vartheta}\psi_{\omega l}(x)\;,
\end{align}
the normalizable solution satisfying the standard boundary condition is
\begin{align}\label{appeq:btzscalarsol}
    \psi^\text{Scalar}_{\omega l}(x) = \frac{(\cosh\left(r_H x \right) )^{\frac{il}{ r_H}}\left(-\sinh \left(r_H x \right) \right)^{\Delta }}{\sqrt{-\tanh(r_H x)}} \, _2F_1\left(a_{\omega l} ,a_{-\omega l};\Delta ;-\sinh^2 \left(r_H x \right) \right) \;,
\end{align}
with
\begin{align}\label{appeq:hyperpara}
    a_{\omega l}=\frac{\Delta}{2}+\frac{i  (-\omega+l )}{2 r_H}\;.
\end{align}


\subsection{Massive vector} \label{app:massvec}

A massive vector on static BTZ is described by the first-order equations 
\begin{align}\label{appeq:1sts1}
    \epsilon_{\alpha\mu\nu}\nabla^\mu A^\nu=\mp m A_\alpha \;. 
\end{align}
Here we take the following convention for the Levi-Civita symbol
\begin{align}
	\epsilon^{\mu\nu\lambda}\equiv \frac{\tilde{\epsilon}^{\mu\nu\lambda}}{\sqrt{-g}} \; , \qquad \tilde{\epsilon}^{-+\vartheta} = -\tilde{\epsilon}_{-+\vartheta} =1 \; .
\end{align} 
The solutions to each of the $\pm$-equations \eqref{appeq:1sts1} furnish an irreducible representation of $SO(2,2)$. In a parity-invariant theory, both sets of solutions should be included. It is straightforward to show that the solutions to each equation satisfy the Fierz-Pauli system
\begin{align}\label{appeq:btzFPs1}
	\left(-\nabla^2+\Delta(\Delta-2)-1 \right)A_{\mu} =0\;, \qquad \nabla^\lambda A_{\lambda } =0 \; ,
\end{align}
where $\Delta=1+m$. We will focus on the components $A_\pm$, which uniquely determine $A_\vartheta$ through the $\vartheta$-component of \eqref{appeq:1sts1}. Working out the $\pm$-components of \eqref{appeq:1sts1} and using the transversality condition \eqref{appeq:btzFPs1}, one finds that $A_\pm$ satisfy
\begin{align}\label{appeq:btzApmeq}
    \left[-\nabla^2_S +(\Delta-1)^2+\tanh^2 (r_H x) -\frac{2}{r_H}\tanh (r_H x) \partial_x \right]A_\pm=0\;,
\end{align}
where $\nabla_S^2$ is the scalar Laplacian:
\begin{equation}\label{appeq:btzsolap}
	\nabla_S^2 \equiv \frac{1}{\sqrt{-g}}\partial_\mu \left( \sqrt{-g}\, \partial^\mu\right)=\frac{\sinh^2 (r_H x)}{r_H^2} (-\partial_t ^2 + \partial_x^2) - \frac{\tanh(r_H x)}{r_H} \partial_x + \frac{\tanh^2(r_H x)}{r_H^2} \partial_\vartheta^2 \;.
\end{equation}
If we further define
\begin{align}\label{appeq:btzs1rec}
    A_\pm = \frac{\bar A_\pm}{\sqrt{-\tanh(r_H x)}} \; ,
\end{align}
we have
\begin{equation}\label{appeq:btzveceq}
	\left[ - \partial_t^2 + \partial_x^2 + \frac{r_H^2}{\sinh^2(2 r_H x)} - \frac{(\Delta-1)^2 r_H^2}{\sinh^2 (r_Hx)} + \frac{\partial_\vartheta^2}{\cosh^2(r_H x)}\right] \bar A_\pm = 0\;.
\end{equation}
These take exactly the same form as the scalar equation \eqref{appeq:btzscalar}. Note that in the near-horizon limit $x \rightarrow -\infty$, \eqref{appeq:btzveceq} reduces to the Rindler-like form \eqref{appeq:rindeqs0}. Since $A_\pm$ are related to $A_t,A_x$ through
\begin{align}\label{appeq:Apmbtz}
    A_\pm  =  \frac{e^{\pm r_H t}}{\tanh(r_H x)} \left(\pm\sinh(r_H x) A_t + \cosh(r_H x) A_x\right)\; ,
\end{align}
normal modes correspond to the ansatz 
\begin{equation} \label{BTZ tilde A ansatz}
	\bar A_\pm = C_{\omega l,\pm} e^{\pm r_H t - i \omega t+i l\vartheta}  \psi_{\omega l, \pm}(x) \;, \qquad A_\vartheta = e^{- i \omega t+i l\vartheta}  \psi_{\omega l, \vartheta}(x)\; .
\end{equation}
Here we have inserted the relative constants $C_{\omega l,\pm}$ to be determined below. Since \eqref{appeq:btzveceq} has exactly the same form as the scalar case \eqref{appeq:btzscalar}, $\psi_{\omega l, \pm}(x)$ are simply given by shifting the scalar solution \eqref{appeq:btzscalarsol} by $\omega \to \omega \pm i r_H$, that is
\begin{align}\label{appeq:btzvecqnmpm}
    \psi_{\omega l, \pm}(x) =&\psi^\text{Scalar}_{\omega\pm i r_H, l}(x)  \;,
\end{align}
where $\psi^\text{Scalar}_{\omega l}(x)$ is defined in \eqref{appeq:btzscalarsol}.

\paragraph{Polarization vectors} 

In the first-order formulation \eqref{appeq:1sts1}, the solutions \eqref{BTZ tilde A ansatz} are not independent. We first consider the $(-)$-branch in \eqref{appeq:1sts1}, which explicitly reads
\begin{align}\label{appeq:1stexp}
	m A_\pm =&\pm\frac{1}{2r_H} \left[  y_\pm^2 \left( \partial_\mp A_\vartheta - \partial_\vartheta A_\mp \right)+(2-y_+ y_-)\left( \partial_\vartheta A_\pm -\partial_\pm A_\vartheta \right)  \right] \;, \nn\\
	m A_\vartheta =& \ 2r_H (1-y_+ y_-)\left( \partial_+ A_- - \partial_- A_+ \right) \;.
\end{align} 
Multiplying the first equation by $y_\pm$ and taking the sum, we have
\begin{align}\label{appeq:pols1inter}
	+m r_H\left(y_+ A_+ +y_- A_- \right) =\frac{1}{r_H}\partial_t A_\vartheta+\partial_\vartheta \left( y_+ A_+ -y_- A_-\right)   \;. 
\end{align}
For the normal mode ansatz \eqref{BTZ tilde A ansatz}, we can replace
\begin{align}
	\partial_t \to - i \omega \;, \qquad \partial_\vartheta \to i l  \; .
\end{align}
Using the $\vartheta$ equation \eqref{appeq:1stexp} in \eqref{appeq:pols1inter}, we then arrive at 
\begin{align}
	 \frac{2i\omega}{m}(1-y_+ y_-)\partial_- A_+ -(m r_H -il) y_+ A_+ = \frac{2i\omega}{m}(1-y_+ y_-)\partial_+ A_- +(m r_H + il) y_- A_- \; .
\end{align}
Plugging in \eqref{BTZ tilde A ansatz} and \eqref{appeq:btzvecqnmpm}, this implies the relation 
\begin{align}
    C^{(-)}_{\omega l,+} (a_{-\omega+i r_H,-l}-1)=- C^{(-)}_{\omega l,-} (a_{\omega+i r_H,+l}-1)
\end{align}
with $a_{\omega l}$ defined in \eqref{appeq:hyperpara}. Here the superscript $(-)$ means that this is associated with the $(-)$-branch \eqref{appeq:1sts1}. Similarly, for the $(+)$-branch we have
\begin{align}
    C^{(+)}_{\omega l,+} (a_{-\omega+i r_H,+l}-1)=- C^{(+)}_{\omega l,-} (a_{\omega+i r_H,-l}-1) \;. 
\end{align}
To summarize, we have determined the two sets of normal modes
\begin{align}\label{appeq:vec1stsol}
    \left(A_+, A_- \right)^{(\mp)}_{\omega l} = e^{ - i \omega t+i l\vartheta} \left(C^{(\mp)}_{\omega l,+} e^{ r_H t } \psi^\text{Scalar}_{\omega+ i r_H, l}(x) , C^{(\mp)}_{\omega l,-}e^{- r_H t } \psi^\text{Scalar}_{\omega- i r_H, l}(x) \right)
\end{align}
satisfying the $\pm$ equations \eqref{appeq:1sts1} respectively.

\subsection{Massive higher spin}\label{app:hSbtz}

Having worked out the warm-up cases of massive scalar and vector, we now study a general spin-$s$ symmetric tensor field $\phi_{\mu_1 \mu_2 \cdots \mu_s}$ with mass $m^2=(\Delta+s-2)(\Delta-s)$, described by either $(\mp)$ set of first-order equations
\begin{align}\label{appeq:1st}
	\epsilon\indices{_{\mu_1}^\alpha^\beta}\nabla_\alpha \phi_{\beta \mu_2 \cdots \mu_s} = \mp M\phi_{\mu_1 \mu_2 \cdots \mu_s} \; , \qquad M \equiv \Delta -1\; .
\end{align}
The solutions to each of the first-order equations \eqref{appeq:1st} satisfy the Fierz-Pauli system \cite{datta_higher_2012}  \begin{align}\label{appeq:btzFP}
    \left(-\nabla^2+\Delta(\Delta-2)-s \right)\phi_{\mu_1 \mu_2 \cdots \mu_s} =0\;, \qquad \nabla^\lambda \phi_{\lambda \mu_1 \mu_2 \cdots \mu_{s-1}} =0 \; , \qquad \phi\indices{^\lambda_{\lambda \mu_1 \mu_2 \cdots \mu_{s-2}}}=0\; .
\end{align}
Solutions to both sets of equations \eqref{appeq:1st} should be included for a parity-invariant theory. 

To proceed, we first note that the action of the Laplacian takes the general form 
\begin{align}\label{appeq:hsbtzlap}
	\nabla^2 \phi_{\mu_1 \mu_2 \cdots \mu_s}
	=& \nabla^2_S\, \phi_{\mu_1 \mu_2 \cdots \mu_s} 
	-\frac{1}{\sqrt{g}}\partial_\alpha \left( \sqrt{g}\, g^{\alpha\lambda}\Gamma^\beta_{\lambda(\mu_1}\right) \phi_{\mu_2 \cdots \mu_s)\beta} 
	 -2\Gamma^\alpha_{\lambda (\mu_1} \nabla^\lambda \phi_{\mu_2 \cdots \mu_{s})\alpha} \nn\\
	 &-g^{\alpha\lambda}\Gamma^\rho_{\alpha\beta}\Gamma^\beta_{\lambda(\mu_1}\phi_{\mu_2 \cdots \mu_s)\rho}
	- \sum_{i\neq j}g^{\alpha\lambda}\Gamma^\beta_{\alpha \mu_i}\Gamma^\rho_{\lambda \mu_j}\phi_{\mu_1  \cdots \hat \mu_i \hat \mu_j \cdots \mu_s\beta \rho}  \;.
\end{align}
In this expression, the symmetrization convention is simply to sum over permutations without extra factors. The summation in the last line has $s(s-1)$ terms. $\nabla^2_S$ is the scalar Laplacian \eqref{appeq:btzsolap}. 

From now on, we use the notation $\phi_{(a)(b)(c)}$ to denote the component of a spin-$s$ symmetric field with $a$ $(+)$-, $b$ $(-)$-, and $c=s-a-b$ $(\vartheta)$-indices, analogous to \eqref{appeq:hsnotation}. For the most part, we will focus on the components with only $\pm$-indices (i.e. those with $c=0$). Solving for these will then uniquely determine the other components through \eqref{appeq:1st}. For these components, after a lengthy calculation we find explicitly (suppressing the $(c=0)$ subscript)
\begin{align}\label{appeq:hsstep1}
	\nabla^2 \phi_{(a)(b)}
	 =& \nabla^2_S\, \phi_{(a)(b)} -2s (1-y_+ y_-) \left( y_+\partial_+ + y_- \partial_- \right) \phi_{(a)(b)} -3s\, \phi_{(a)(b)}  +s^2y_+ y_-\phi_{(a)(b)}  \;. 
\end{align}
In deriving this, we have simplified in \eqref{appeq:hsbtzlap} the second term using
\begin{align}
    \frac{1}{\sqrt{g}}\partial_\alpha \left(\sqrt{g}\,g^{\alpha\lambda}\Gamma^\beta_{\lambda\mu} \right)
    =&
    \begin{cases}
        3 \; , &\beta = \mu=\pm \\
        2\;, &\beta = \mu=\vartheta \\
        0 \;, &\text{otherwise}
    \end{cases} \;,   
  \end{align}
 the third term using the transversality condition \eqref{appeq:btzFP}, the fourth term using
    \begin{align}
    g^{\alpha\lambda} \Gamma^\rho_{\alpha \beta}\Gamma^\beta_{\lambda \mu}
    =&
    \begin{cases}
        2y_+ y_- \; , &\rho = \mu=\pm \\
        y_\pm^2\;, &\rho = \pm \; , \mu=\mp \\
        0 \;, &\text{otherwise}
    \end{cases} \; ,
    \end{align}
and the last term using
    \begin{align}
    g^{\alpha\lambda}\Gamma^\beta_{\alpha \pm}\Gamma^\rho_{\lambda \pm} =&\left(\frac{1}{2} \Gamma^\pm_{\pm \pm}\right)^2 g^{\beta\rho}+ \delta^\beta_\pm \delta^\rho_\pm 2 y_+ y_- +\left( \delta^\beta_+ \delta^\rho_- +\delta^\beta_- \delta^\rho_+\right) \frac{y_\mp^2}{2} \;,\nn\\
	g^{\alpha\lambda}\Gamma^\beta_{\alpha \pm}\Gamma^\rho_{\lambda \mp} =&\frac14 \Gamma^+_{++}\Gamma^-_{--}g^{\beta\rho}+\delta^\beta_+ \delta^\rho_+ \frac{y_+^2}{2}+\delta^\beta_- \delta^\rho_- \frac{y_-^2}{2}+\delta^\beta_\pm \delta^\rho_\mp 2y_+ y_-  \;,
\end{align}
together with the tracelessness condition \eqref{appeq:btzFP}. 

In terms of the variables $t,x$, we can write \eqref{appeq:hsstep1} as
\begin{align}
	\nabla^2 \phi_{(a)(b)}
	 =&\left( \frac{\sinh^2 (r_H x)}{r_H^2} (-\partial_t ^2 + \partial_x^2) - \frac{\tanh(r_H x)}{r_H} \partial_x + \frac{\tanh^2(r_H x)}{r_H^2} \partial_\vartheta^2\right) \phi_{(a)(b)}\nn\\
	 & +2s\frac{\tanh\left(r_H x \right) }{r_H} \partial_x \phi_{(a)(b)} -3s\, \phi_{(a)(b)} +s^2 \sech^2 \left(r_H x \right) \phi_{(a)(b)} \;. 
\end{align}
Finally, rescaling
\begin{align}\label{appeq:btzhsrec}
    \phi_{(a)(b)} = (-\tanh(r_H x))^{\frac{1}{2}-s}\bar \phi_{(a)(b)} \; ,
\end{align}
we find that the second-order equations \eqref{appeq:btzFP} for these components are reduced to 
\begin{equation}\label{appeq:btzspineq}
	\left[ - \partial_t^2 + \partial_x^2 + \frac{r_H^2}{\sinh^2(2 r_H x)} - \frac{(\Delta-1)^2 r_H^2}{\sinh^2 (r_Hx)} + \frac{\partial_\vartheta^2}{\cosh^2(r_H x)}\right] \bar \phi_{(a)(b)} = 0\;.
\end{equation}
Therefore, we have a set of decoupled equations that take the scalar form \eqref{appeq:btzscalar}. Dictated by the relations between $\pm$- and $t,x$-components, normal modes correspond to the ansatz 
\begin{equation} \label{BTZ tilde hs ansatz}
	\bar \phi_{(a)(b)(c)} = C_{\omega l,(a)(b)(c)}\, e^{ (a-b) r_H t - i \omega t+i l\vartheta}  \psi_{\omega l, (a)(b)(c)}(x) \; .
\end{equation}
This is true even for $c\neq 0$. Here $C_{\omega l,(a)(b)(c)}$ are polarization constants to be determined. 

Since \eqref{appeq:btzspineq} has exactly the same form as the scalar case \eqref{appeq:btzscalar}, $\psi_{\omega l, (a)(b)}(x)$ are simply given by shifting the scalar solution \eqref{appeq:btzscalarsol} by $\omega \to \omega + i(a-b) r_H$, that is
\begin{align}\label{appeq:btzhsqnmpm}
    \psi_{\omega l, (a)(b)}(x) =\psi^\text{Scalar}_{\omega+ i (a-b) r_H, l}(x) \;,
\end{align}
with $\psi^\text{Scalar}_{\omega l}(x)$ defined in \eqref{appeq:btzscalarsol}.

\paragraph{Polarization tensors}

In the first-order formulation \eqref{appeq:1st}, the solutions \eqref{appeq:btzhsqnmpm} are not independent. Analogous to the vector case, we multiply the $(a)(b)(c)$- and $(a-1)(b+1)(c)$-components of \eqref{appeq:1st} by $y_+$ and $y_-$ respectively and take their sum, which leads to
\begin{align}\label{appeq:hspolin}
    \pm M r_H\left(y_+ \phi_{(a)(b)(c)} +y_- \phi_{(a-1)(b+1)(c)} \right)  =\ &\frac{1}{r_H}\partial_t \phi_{(a-1)(b)(c+1)} +\partial_\vartheta \left( y_+ \phi_{(a)(b)(c)} -y_- \phi_{(a-1)(b+1)(c)}\right)  \nn\\
    &+ c \, r_H^2 \left(y_+^2 \phi_{(a+1)(b)(c-1)} -y_-^2 \phi_{(a-1)(b+2)(c-1)}\right) \; .
\end{align}
Here the upper (lower) sign corresponds to the $-$ (+)-branch \eqref{appeq:1st}. For the normal mode ansatz \eqref{BTZ tilde hs ansatz}, we can replace
\begin{align}
    \partial_t \to  - i \omega +(a-1-b) r_H \; , \qquad \partial_\vartheta \to i l \;.
\end{align}
For $c=0$, using also the $(a-1)(b)(c+1)$-component of \eqref{appeq:1st}, \eqref{appeq:hspolin} leads to the relation
\begin{align}
	 &\frac{2i(\omega+i(a-1-b)r_H)}{M}\left((1-y_+ y_-)\partial_- \phi_{(a)(b)}-\frac{s-1}{2}y_+\phi_{(a)(b)}\right) -(M r_H \mp il) y_+ \phi_{(a)(b)} \nn\\
	 =& \frac{2i(\omega+i(a-1-b)r_H)}{M}\left((1-y_+ y_-)\partial_+ \phi_{(a-1)(b+1)}-\frac{s-1}{2}y_-\phi_{(a-1)(b+1)}\right) \nn\\
	 &+(M r_H \pm il) y_- \phi_{(a-1)(b+1)} \; .
\end{align}
Substituting \eqref{BTZ tilde hs ansatz} and \eqref{appeq:btzhsqnmpm}, we arrive at the recursion relation
\begin{align}\label{appeq:btzhsrecur}
    C^{(\mp)}_{\omega l,(a)(b)} (a_{-\omega-i(a-b-2) r_H,{\mp} l}-1)=- C^{(\mp)}_{\omega l,(a-1)(b+1)} (a_{\omega+i (a-b)r_H,\pm l}-1) \;,
\end{align}
with $a_{\omega l}$ defined in \eqref{appeq:hyperpara}. Here $(\mp)$ means that this is associated with the $\mp$-branch \eqref{appeq:1st}. Using \eqref{appeq:btzhsrecur}, it is straightforward to derive the relation \eqref{eq:polhsbtz}.

 \bibliographystyle{utphys}
 \bibliography{BHChar}

\end{document}